\documentclass[11pt,usletter]{article}
\pdfoutput=1
\usepackage{jheppub}
\usepackage[utf8]{inputenc}
\usepackage{epsfig,multicol}
\usepackage[sort&compress,numbers]{natbib}
\usepackage{graphicx}
\newcommand{\beq}{\begin{equation}}
\newcommand{\eeq}{\end{equation}}
\newcommand{\beqa}{\begin{eqnarray}}
\newcommand{\eeqa}{\end{eqnarray}}

\newcommand{\brem}{bremsstrahlung\ }

\title{A Projective Phase Space Generator for 
Hadronic Vector Boson Plus One Jet Production}

\author[a]{Tinghua Chen}
\author[a]{Terrance M.~Figy}
\affiliation[a]{Department of Mathematics, Statistics, and Physics,
Wichita State University, Wichita, Kansas, USA}
\author[b]{and Walter T.~Giele}
\affiliation[b]{Theory Group, Fermilab, Batavia, USA}
\emailAdd{txchen2@shockers.wichita.edu}
\emailAdd{terrance.figy@wichita.edu}
\emailAdd{giele@fnal.gov}

\abstract{
In this paper we use our previously developed projective phase space generator 
for the calculation of the hadronic production of a vector boson with 
one additional jet at Next-to-Leading Order.
The projective phase space generator allows us to make physical predictions
in novel ways, speeding up both evaluation time and attainable accuracy.  

For the numerical evaluation, we explore a computational model
which combines the use of both multi-threading and
distributed resources through the use of grid or cloud computing
without depending on local institutional computer availability.
The projective phase space method is well suited for
this approach and gives through the use of cloud computing
instant access to a large pool of resources.
}

\preprint{preprint FERMILAB-PUB-19-300-T}
\arxivnumber{1907.03893}
\notoc
\begin{document}
\maketitle

\newpage

\section{Introduction}

The use of projective phase space methods opens up new
ways to perform phase space integration at higher orders.
The main advantage is the factorization of phase space 
into a physical Born phase space 
and a marginalized \brem phase space.
Any observable is defined on the Born phase space,
while the \brem phase space is integrated over with the constraint
that the observables are unaltered.
As a result, the radiative corrections
are given as $K$-factors to the observable defined from the  Born
kinematics. The remaining challenge is connecting the Born defined observables 
with the experimental observables. 

When this topic is discussed in the literature,
it is typically from the viewpoint of the Matrix Element 
Method~\cite{Alwall:2010cq,Giele:2011tm,Campbell:2012cz,Campbell:2012ct,Campbell:2013hz,Martini:2015fsa,Martini:2017ydu,Kraus:2019qoq}.
We follow a different approach as outlined in Ref.~\cite{Figy:2018imt}
where we view the topic as a simple phase space integration technique and
consequently an issue of defining observables 
instead of the more restricted subject of Matrix Element Methods.

While constructing a projective phase space integration method
is not particularly complicated, it becomes more involved
when one wants to stay connected to a realistic experimental
setting. 
Given a hadronic event, one can project
this event to an unique Born event with a Leading Order (LO) weight
using a jet algorithm. Any jet algorithm can be viewed as a projector
of the hadronic event onto the partonic Born hard scattering event
as it marginalizes out the particle content, 
thereby reducing the phase space
dimensionality to the Born phase space dimensionality. 
Several issues will immediately arise, which will constrain a jet algorithm
usable for a projective phase space integration method. 
First, a jet algorithm does not necessarily 
construct a massless 4-vector for the jet axis. Secondly, the reconstructed
objects in the event are not necessarily balanced in transverse energy
due to initial state radiation. The latter issue is specific for hadron colliders.
Both these issues can cause problems for a perturbative prediction.
An infrared stable jet algorithm  should take these issues
into consideration as the lowest order prediction has both massless
jets and the objects in the 
event are strictly balanced in transverse momentum. 
The perturbative issues can be handled rigorously while the
non-perturbative issues resulting from
hadronization and the beam/proton remnants are the root
of some fundamental issues that are more difficult to assess.

The addition of \brem to a Born jet event makes
changes to the event using traditional jet algorithms. 
One produces either an additional 
jet through wide angle hard branching, generate a
$p_T$-imbalance in the reconstructed objects through initial state radiation or
obtain a massive jet through soft/collinear final state branchings. 
The latter two types
of branchings need to be integrated over, that is the radiation
needs to be inclusive in order to obtain an infrared safe higher order 
prediction given the fixed Born event. Traditionally, one defines the observables
using the reconstructed objects which in turn are constructed from events generated
during a numerical integration
over the \brem phase space. This requires explicit finite width histogramming
to combine the \brem events with the virtual events to obtain the infrared safe prediction
for the observable.
Alternatively, one can specify the objects in advance and
integrate over the radiation contributing to
the fixed objects. This has the advantage one can define any observable
using these inclusive objects without the need to consider infrared stability
as virtual corrections and \brem have been properly combined 
at the object level without referring to either observable or explicit histogramming.

From the perspective of fixed higher order correction the inclusion
of the \brem directly into the 
object definition leads naturally to projective phase space techniques
and the factorization into a Born phase space times a \brem phase space. 
This results in
many advantages over traditional phase space Monte Carlo integration
techniques.  
While the integration over the jet mass is conceptually easy to
understand, the treatment of the initial state radiation is more complicated.
In Ref.~\cite{Figy:2018imt} we developed a forward branching phase space
(FBPS) generator which leaves key jet observables invariant with respect to the
Born event. As a consequence any observable ${\cal O}$ constructed from
the final state objects and hence depending only on the Born kinematics
can be calculated exclusively. That is, a Higher Order (HO) prediction
of observable ${\cal O}$ is given by a $K$-factor times the LO prediction,
where $K({\cal O})$ can be calculated order by order in perturbation
theory without altering the observable ${\cal O}$. Note that these
methods hark back to pre-Monte Carlo analytic phase space
integration techniques (see e.g.~\cite{Ellis:1990ek}), 
but now implemented fully numerically offering all the flexibility
coming with such an approach. 

In Sec.~2 we discuss the definition of observables through an inclusive and
exclusive jet algorithm at fixed order. We also discuss how to connect
to realistic observables by including parton showers, hadronization 
and beam remnants. These methods are worked out in Sec.~4 for fixed order
and in Sec.~5 for hadronized observables. Sec.~3 studies the slicing parameter 
dependence as we have not yet worked out a subtraction method compatible with 
the FBPS method. In Sec.~6 the ideas introduced in this paper are summarized.

\section{Observables}

At hadron colliders observables are  usually built
out of the boost invariant kinematic quantities of the reconstructed
objects, namely transverse momentum, rapidity and azimuthal angle.
Using these kinematic quantities to build observables enhances
the infrared stability of these observables,
while depending on other kinematic quantities which are not present at LO 
such as jet mass can cause infrared instabilities because they will in
general constrain \brem radiation.

The FBPS formalism projects events onto the Born kinematics and is
constructed to predict observables using any infrared safe jet
algorithm where the only restriction is that two clusters are
combined sequentially by adding the respective 4-vectors of the two clusters.
The projection is designed to leave the key kinematic quantities 
invariant. That is, the transverse momentum vector and rapidities of the observables
are unaltered. These kinematic quantities completely fix the projected Born phase 
space point to a single event. 
Specifically, for the process $PP\rightarrow V + n$ jets,
the FBPS formalism calculates 
the $K$-factor order-by-order in pQCD for the maximum 
exclusive observable ${\cal O}_{\mbox{\tiny MAX}}$
\beqa\label{fbpsO}
\left.\frac{d\sigma}{d{\cal O}_{\mbox{\tiny MAX}}}\right\rfloor_{HO}
&=&K({\cal O}_{\mbox{\tiny MAX}})\times
\left.\frac{d\sigma}{d{\cal O}_{\mbox{\tiny MAX}}}\right\rfloor_{LO}\nonumber \\
d\,{\cal O}_{\mbox{\tiny MAX}}&=&d\,\eta_V\times\prod_{i=1}^n\left(d\,\vec p_T^{(i)} d\,\eta_J^{(i)}\right)\ ,
\eeqa
point-by-point in phase space. (Note that we suppressed sums over partonic flavors for simplicity.) 
The transverse momentum vector and rapidity of jet $i$ are given by 
$\vec p_T^{(i)}$ and $\eta_J^{(i)}$, respectively. The vector boson rapidity is given by $\eta_{V}$.
Composite observables can be built from the maximal exclusive differential cross section.
A noteworthy example is the value of the transverse momentum of the vector boson. At Born level this
is constrained by momentum conservation $\vec p_T^V=-\sum_i \vec p_T^{(i)}$. At higher
orders this becomes more complicated as not all final state partons are necessarily
clustered into a jet $\vec p_T^V+\vec p_T^{\mbox{\tiny UNC}}=-\sum_i \vec p_T^{(i)}$ where
$\vec p_T^{\mbox{\tiny UNC}}$ is the transverse momentum vector of the summed momenta of the
unclustered partons. For the needed projection onto Born any $\vec p_T^{\mbox{\tiny UNC}}$
generated by the jet algorithm
needs to be added to the vector boson, while the longitudinal vector boson momentum is re-scaled
to leave the vector boson mass invariant. Finally, we can swap the vector boson momentum
with one of the jet momenta in Eq.~\ref{fbpsO} if needed for constructing a composite
observable containing the vector boson, 
i.e. $(\vec p_T^V,\eta_V)\leftrightarrow (\vec p_T^{(k)},\eta_J^{(k)})$.
In practice the $\vec p_T$-imbalance is easily avoided at parton level.

We use an inclusive partonic jet algorithm which terminates
when a set number of $n$ clusters are obtained. No beam jet is constructed 
in this jet algorithm and all partons
are clustered into the jet(s), i.e. $\vec p_T^V+\sum_i \vec p_T^{(i)}=0$. For a partonic
final state this jet algorithm works well and simplifies the Monte Carlo implementation
of the FBPS formalism significantly.
An exclusive partonic jet algorithm can be simply defined by adding a $(n+1)$-jet 
veto on the inclusive jet algorithm. This means the FBPS will integrate over the jet cones and
initial state radiation, while excluding the $(n+1)$-jet region of phase space. 
We can e.g. choose an anti-$k_T$ jet algorithm~\cite{Cacciari:2008gp}. 
However, it will project to a perfectly balanced Born event. That is all partons
are clustered into one of the jets. It is worth noting that we integrate the 
\brem partons exactly over the $n$-jet phase space defined by the anti-$k_T$ algorithm.
Only the object reconstruction differs between partonic and hadronic as for the latter
we can have unclustered hadrons.

Before being able to compare with experimental data we have to consider hadronization
and more relevant the proton/beam remnants which themselves carry a color charge. This
forces a hadronic jet algorithm to construct a beam jet to ensure
the proton remnant hadronic contribution into the hard scattering content
is minimized. 
By using the partonic jet algorithms defined above we can still connect to a realistic
experimental environment through the inclusion of a shower 
Monte Carlo~\cite{Bahr:2008pv,Gleisberg:2008ta,Sjostrand:2014zea}. 
In the FBPS formalism all partonic 
final states contributing to a single (multi-) jet event with a specified jet final state
are integrated over, resulting in Eq.~\ref{fbpsO}.
This means we can shower the jets using any LO matching formalism initiated from the
projected Born event. The shower will repopulate the regions of \brem phase
space which were integrated out on the partonic level.

The result is for the inclusive partonic jet algorithm of
a fully exclusive hadronic final state 
where all radiative corrections are contained in the $K$-factor
weight. Because showers are unitary, the $K$-factor will not be altered. 
Any experimental jet algorithm can be applied to the now fully exclusive
hadronic final state. This will recast the partonic jets into the hadronic jets and
the partonic Born phase space point will be smeared around this point.
The result is that an observable does not necessarily contributes to a single bin, but can
spread over multiple bins. Yet the weight of this distribution is still given
by Eq.~\ref{fbpsO} and one can simply sum over the generated list of (possibly unit-weight)
LO events, re-weighted by the $K$-factor.
The only requirement of the shower is that it is unitary. In this scenario we rely on the shower
MC to generate the hard branchings to give us additional jets. 
We will, in Sec. 5 use the {\tt VINCIA} plugin~\cite{Giele:2007di,Giele:2011cb,Fischer:2016vfv} 
to {\tt PYTHIA}~\cite{Sjostrand:2006za,Sjostrand:2014zea}
for this purpose as it is Matrix Element (ME) corrected. 
Without a ME improved shower for the first branching(s) this is not expected to work well.

Without a ME corrected shower we can improve the hadronic event generation by employing
a MLM matching scheme~\cite{Mangano:2006rw}. This will truncate the wide angle radiation
generated by the parton shower from the NLO inclusive $Z+n$-jet sample 
and hence dynamically reducing the size of the overall $K$-factor.
The vetoed inclusive $(n+1)$-jet phase space region will be repopulated by showered LO 
$Z+(n+1)$-jet events, thereby negating the necessity of the ME corrected shower.
In this paper we will not explore this particular option any further.

We see that by using a projective phase space integration method we can use a theoretical partonic
jet algorithm to project on the Born phase space and yet through the use of a unitary
shower Monte Carlo recast the events as fully hadronic events which can be reconstructed
using any experimental jet algorithm and apply detector effects. The $K$-factor of the
fully hadronic event can be calculated using the FBPS formalism order-by-order in
perturbative QCD.
Note that a Monte Carlo integration over the showered and $K$-factor re-weighted LO events is
needed as the events are smeared out.

As a first exploration of these concepts we look at $PP\rightarrow Z+1$ jet where
we use our partonic jet algorithms including the recasting to hadronic jet algorithms
using  a shower Monte Carlo . 
The $K$-factor for this process is calculated up to Next-to-Leading order (NLO) ${\cal O}(\alpha_S^2)$. 

\section{Validation of the Slicing Method}
\begin{figure}[t]
\begin{center}
\includegraphics[width=7.7cm]{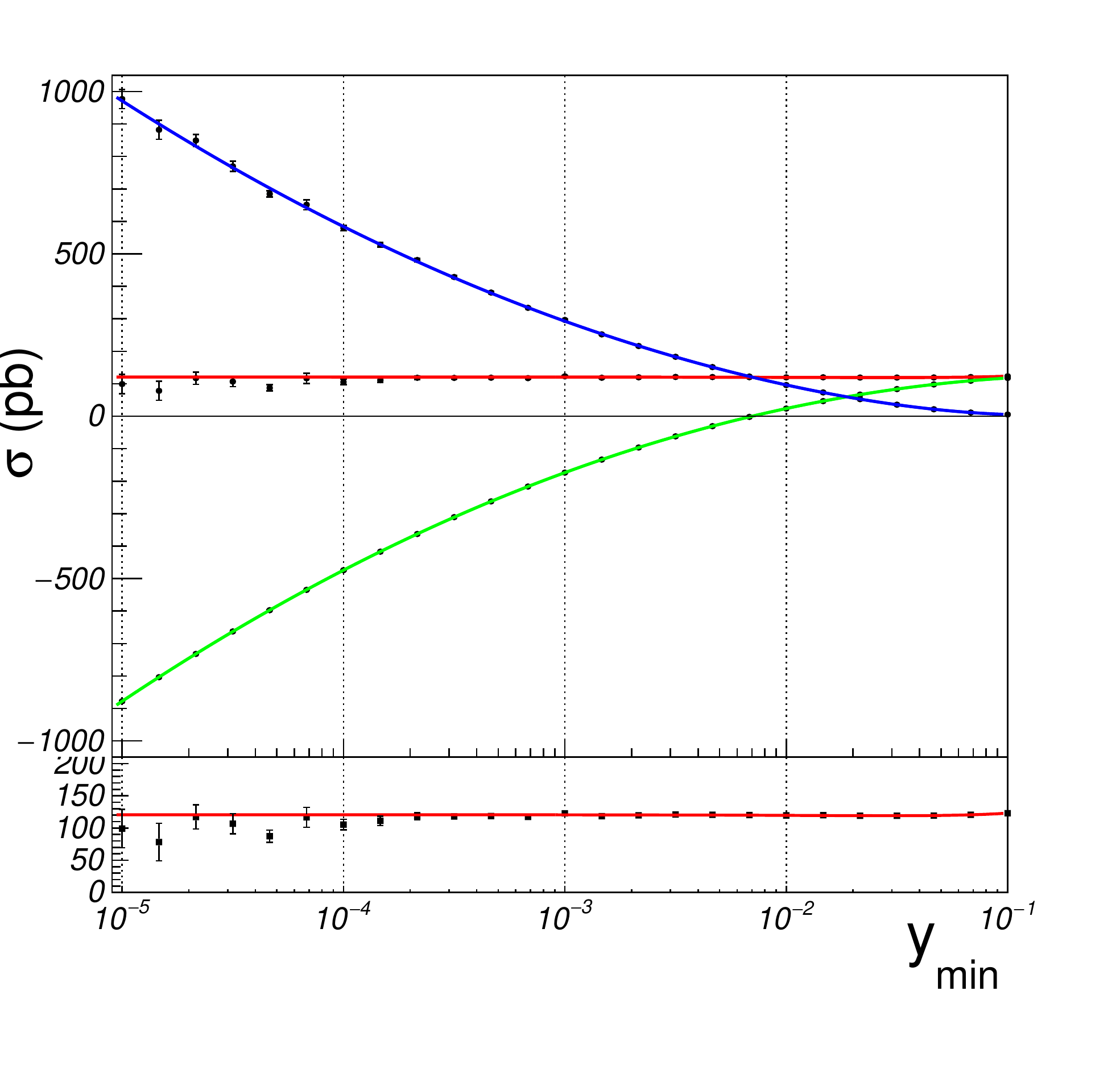}
\includegraphics[width=7.7cm]{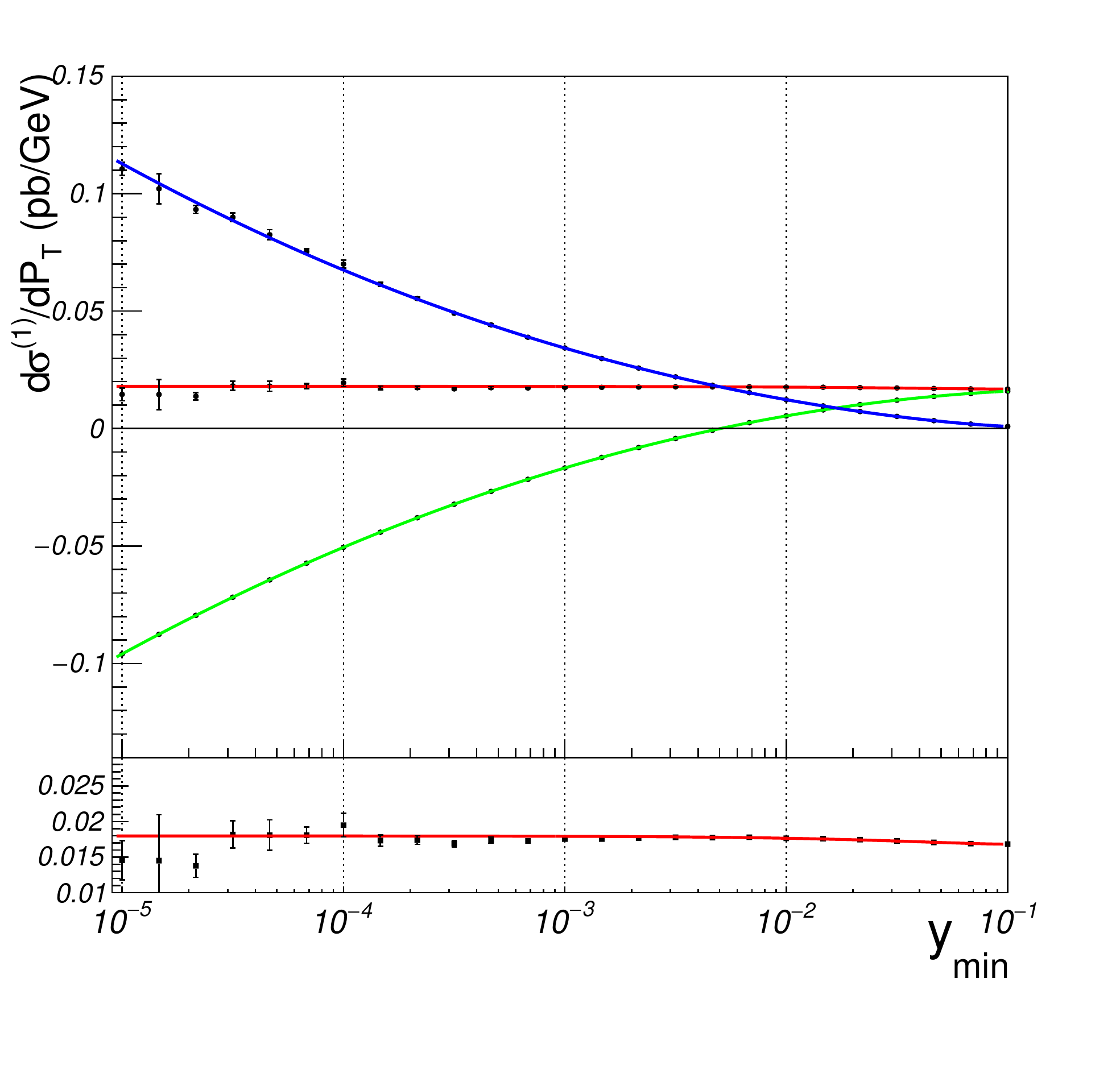}
\includegraphics[width=7.7cm]{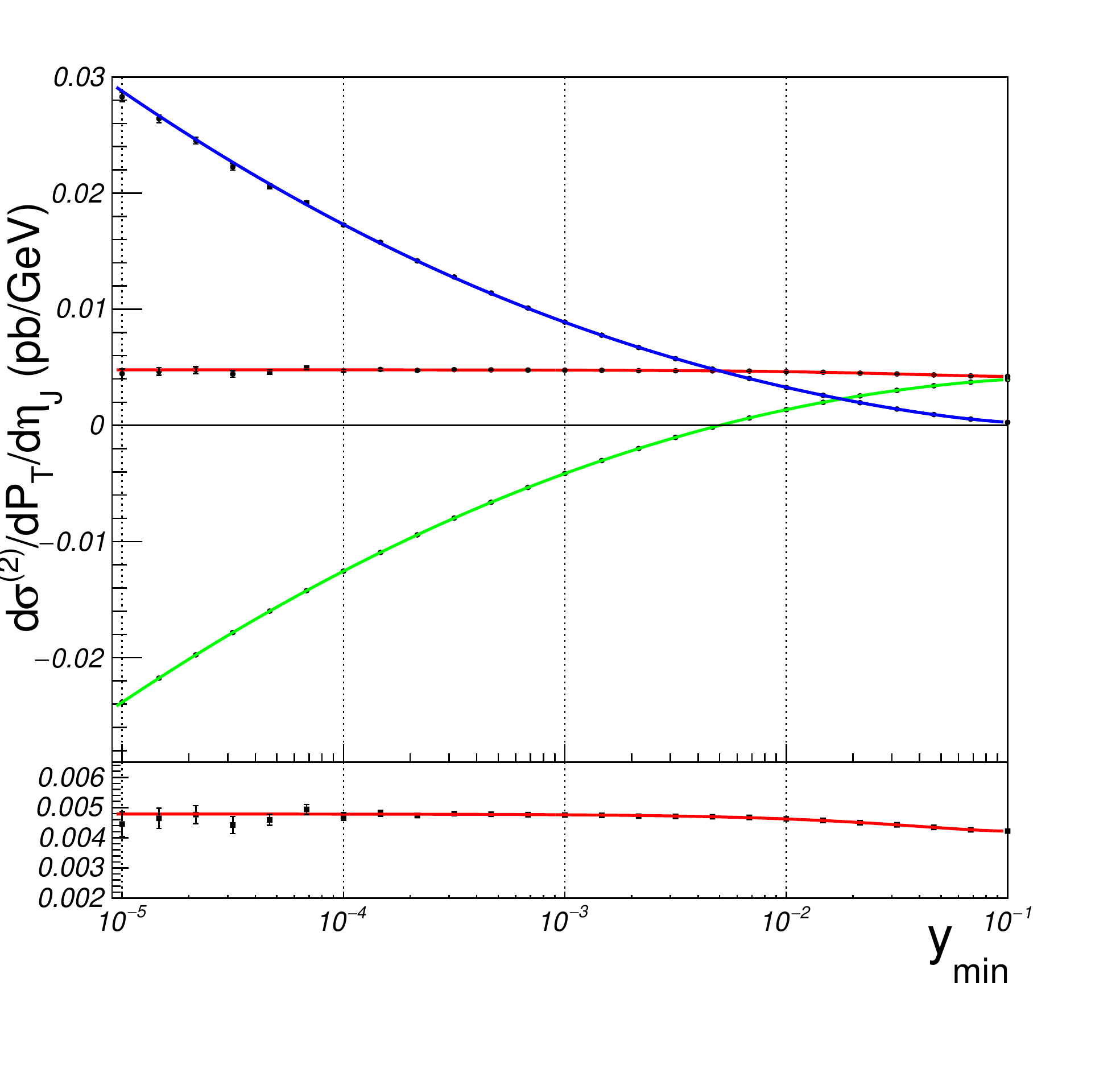}
\includegraphics[width=7.7cm]{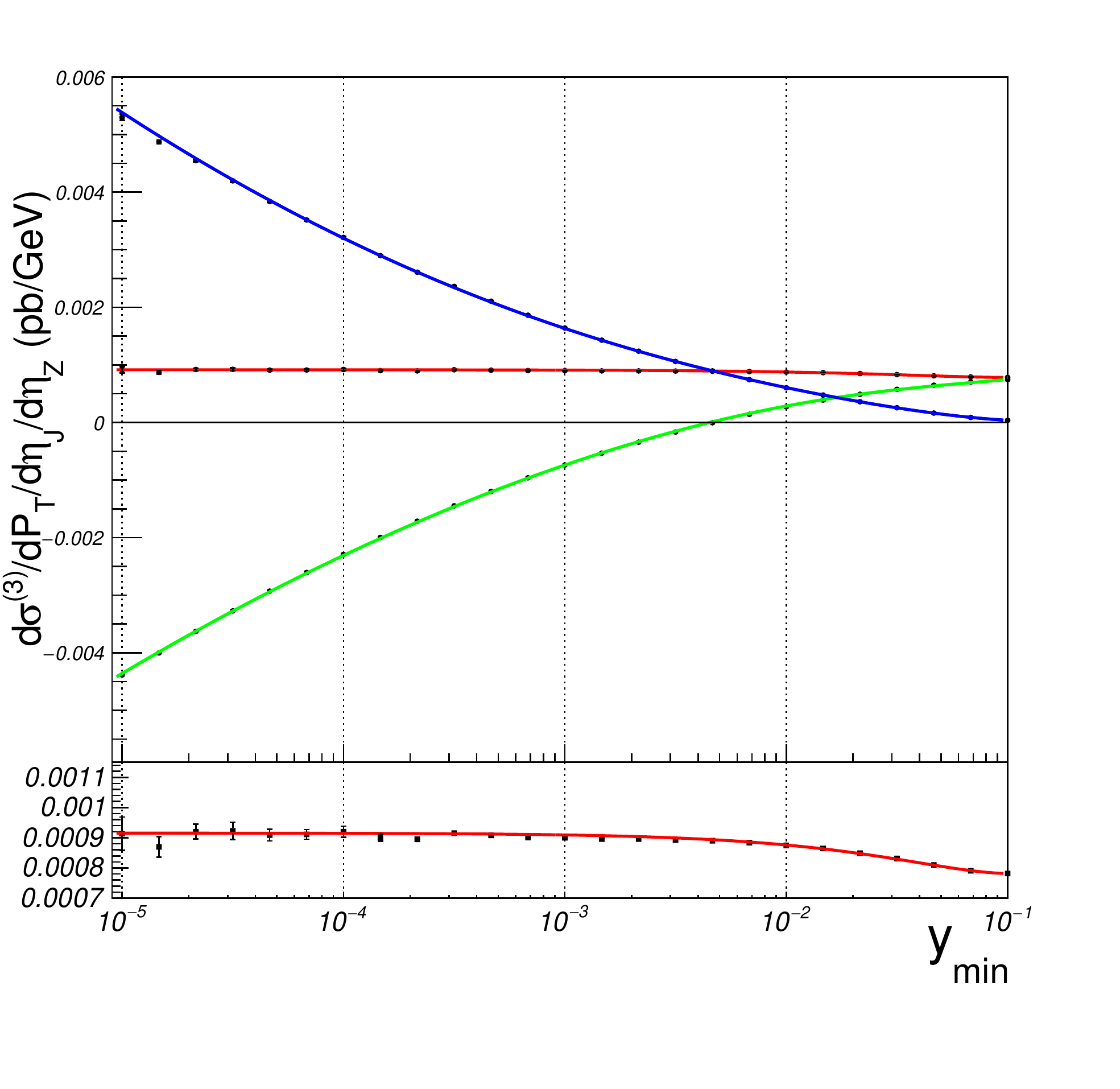}
\end{center}
\caption{The $y_{\mbox{\tiny min}}$ dependence of the inclusive $PP\rightarrow Z+1$ jet production 
at $\sqrt{S}$=14 TeV. Details are specified in text.
The blue curve is the \brem contribution, the green curves
is the virtual contribution and the sum of the two contributions
is represented by the red curve.
The inclusive cross section is shown top left.
The single differential distribution
$d\sigma^{(1)}/d p_T^{\mbox{\tiny jet}}$ at $p_T^{\mbox{\tiny jet}}=250$ GeV 
is shown top right.
The double differential distribution
$d\sigma^{(2)}/d p_T^{\mbox{\tiny jet}}/d \eta_{\mbox{\tiny jet}}$ 
at $p_T^{\mbox{\tiny jet}}=250$ GeV and $\eta_{\mbox{\tiny jet}}=-0.75$ 
is shown bottom left.
Finally, on the bottom right is the triple differential distribution
$d\sigma^{(3)}/d p_T^{\mbox{\tiny jet}}/d \eta_{\mbox{\tiny jet}}/d \eta_Z$ 
at $p_T^{\mbox{\tiny jet}}=250$ GeV, $\eta_{\mbox{\tiny jet}}=-0.75$ and
$\eta_Z=0.51$.
}
\label{ymindep}
\end{figure}

While the FBPS method does not require slicing, a compatible subtraction method 
needs to be developed. For now we use a slicing method which fits well within 
a FBPS approach. This method can be used without any modifications.
We will take $PP\rightarrow V+1$ jet production at NLO using a slicing method as implemented
in the DYRAD parton level MC~\cite{Giele:1993dj} for all numerical results in this paper.

As a validation of the method and behavior with respect to the slicing parameter
we look at four observables with increasing final state exclusivity. For all
the MC runs we use {\tt VEGAS}~\cite{Lepage:1977sw} 
as implemented in the CUBA library~\cite{Hahn:2004fe}.
The use of the CUBA library extends the DYRAD MC to include multi-threading,
speeding up the run-time considerably depending on the available threads. The results
for the dependence on the slicing parameter of the four observables
are shown in Fig.~\ref{ymindep}.
We use a collider energy of $\sqrt{S}=14$~TeV, the renormalization and factorization 
scales set to $\mu_R=\mu_F=\sqrt{\hat s}$ with $\hat s$ defined as the partonic
collision energy, and the CT14nlo PDF set~\cite{Buckley:2014ana}. 
Shown are the real and virtual contributions together with the sum of these two
contributions. This combined result is fitted to the expected leading behavior
of the slicing parameter
$y_{\mbox{\tiny MIN}}=s_{\mbox{\tiny MIN}}/\hat s$ for observable ${\cal O}$
\beq
{\cal O}(y_{\mbox{\tiny MIN}})={\cal O}(0)\times\left(1+\left(a_1\times y_{\mbox{\tiny MIN}}
+a_2\times y_{\mbox{\tiny MIN}}^2\right)\times\log\left(y_{\mbox{\tiny MIN}}\right)
+\cdots\right)\ ,
\eeq
where $s_{\mbox{\tiny MIN}}$
is the minimal invariant mass between two partons during the \brem phase
space integration.
For the MC run to calculate the virtual and \brem cross sections
we selected a target relative accuracy of 0.1\% with an event cap
of 10,000,000.

The first observable we consider is the inclusive cross section 
$\sigma$ where the transverse jet momentum $p_T^{\mbox{\tiny JET}}>50$ GeV and
the jet rapidity
$\eta_{\mbox{\tiny JET}}<3$. This is a traditional (3+3)-dimensional phase space 
integration (3 dimensions for the Born phase space plus 3 dimensions for the \brem
phase space). The result is summarized in
\beqa
\sigma_{\mbox{\tiny LO}} &=&(1.028\pm 0.001)\times 10^2\ \mbox{pb} \nonumber \\
{\cal O}(0)=\sigma_{\mbox{\tiny NLO}}&=&(1.207\pm 0.003)\times 10^2\ \mbox{pb}\ ,
\eeqa
where the fitting parameters are given $a_1=0.21\pm 0.01$ and $a_2=-3.0\pm 0.2$.

The second observable is the single differential $p_T^{\mbox{\tiny JET}}$
cross section at $p_T^{\mbox{\tiny JET}}=250$ GeV and $\eta_{\mbox{\tiny JET}}<3$.
This reduces the phase space dimensionality to (2+3) 
(2 dimensions for the Born phase space plus 3 dimensions for the \brem
phase space), resulting in
\beqa
\left.\frac{d\sigma^{(1)}_{\mbox{\tiny LO}}}{d p_T^{\mbox{\tiny jet}}}
\right\rfloor_{p_T^{\mbox{\tiny jet}}=250\ \mbox{\tiny GeV}}
&=&(1.318\pm 0.001)\times 10^{-2}\ \mbox{pb/GeV}\nonumber \\
{\cal O}(0)=\left.\frac{d\sigma^{(1)}_{\mbox{\tiny NLO}}}{d p_T^{\mbox{\tiny jet}}}
\right\rfloor_{p_T^{\mbox{\tiny jet}}=250\ \mbox{\tiny GeV}}
&=&(1.797\pm 0.004)\times 10^{-2}\ \mbox{pb/GeV}\ ,
\eeqa
where the fitting parameters are given by $a_1=0.41\pm 0.03$ and  $a_2=-1.2\pm 0.2$.

The dimensionality of the third observable is further reduced to (1+3).
This observable is the double differential cross section at the point 
$p_T^{\mbox{\tiny JET}}=250$ GeV and $\eta_{\mbox{\tiny JET}}=-0.75$ giving
\beqa
\left.\frac{d\sigma^{(2)}_{\mbox{\tiny LO}}}{d p_T^{\mbox{\tiny jet}}d\eta_{\mbox{\tiny jet}}}
\right\rfloor_{p_T^{\mbox{\tiny jet}}=250\ \mbox{\tiny GeV},\, \eta_{\mbox{\tiny jet}}=-0.75}
&=&(3.296\pm 0.003)\times 10^{-3}\ \mbox{pb/GeV} \nonumber \\
{\cal O}(0)=\left.\frac{d\sigma^{(2)}_{\mbox{\tiny NLO}}}{d p_T^{\mbox{\tiny jet}}d\eta_{\mbox{\tiny jet}}}
\right\rfloor_{p_T^{\mbox{\tiny jet}}=250\ \mbox{\tiny GeV},\, \eta_{\mbox{\tiny jet}}=-0.75}
&=&(4.785\pm 0.006)\times 10^{-3}\ \mbox{pb/GeV}\ ,
\eeqa
where the fitting parameters are given by $a_1=0.78\pm 0.02$ and $a_2=-2.5\pm 0.1$.

Finally, we restrict the Born kinematics to a single event (apart from the
trivial azimuthal event angle). This leaves the (0+3) variables describing the
\brem particle left to integrate over. The result is given by,
\beqa
\left.\frac{d\sigma^{(3)}_{\mbox{\tiny LO}}}{d p_T^{\mbox{\tiny jet}}d\eta_{\mbox{\tiny jet}}d\eta_Z}
\right\rfloor_{p_T^{\mbox{\tiny jet}}=250\ \mbox{\tiny GeV},\, \eta_{\mbox{\tiny jet}}=-0.75,\, \eta_Z=0.51}
&=&(6.169\pm 0.000)\times 10^{-4}\ \mbox{pb/GeV} \nonumber \\
{\cal O}(0)=\left.\frac{d\sigma^{(3)}_{\mbox{\tiny NLO}}}{d p_T^{\mbox{\tiny jet}}d\eta_{\mbox{\tiny jet}}d\eta_Z}
\right\rfloor_{p_T^{\mbox{\tiny jet}}=250\ \mbox{\tiny GeV},\, \eta_{\mbox{\tiny jet}}=-0.75,\, \eta_Z=0.51}
&=&(9.154\pm 0.007)\times 10^{-4}\ \mbox{pb/GeV}\ , \nonumber \\
\eeqa
where the fitting parameters are given by $a_1=0.97\pm 0.01$ and $a_2=-3.3\pm 0.4$.

Given the accuracy of the MC runs (a 0.1\% statistical uncertainty on the virtual
and \brem MC integrations), we choose as a default the value of $y_{\mbox{\tiny MIN}}=10^{-3}$
for the remainder of the paper. Note that the more exclusive observables have
a stronger dependence on $y_{\mbox{\tiny MIN}}$. The value chosen 
for the slicing parameter should be safe for any observable at this accuracy.

As an alternative to {\tt VEGAS} we also tried {\tt SUAVE}~\cite{Hahn:2004fe}. For $y_{\mbox{\tiny MIN}}>10^{-4}$
no statistically significant differences were observed. Below this value {\tt SUAVE}
underestimates the value of the cross section. We therefore will use {\tt VEGAS}
as implemented in CUBA for the remainder of the paper.

\section{Theoretical Predictions}
\begin{figure}[t]
\begin{center}
\includegraphics[width=7.7cm]{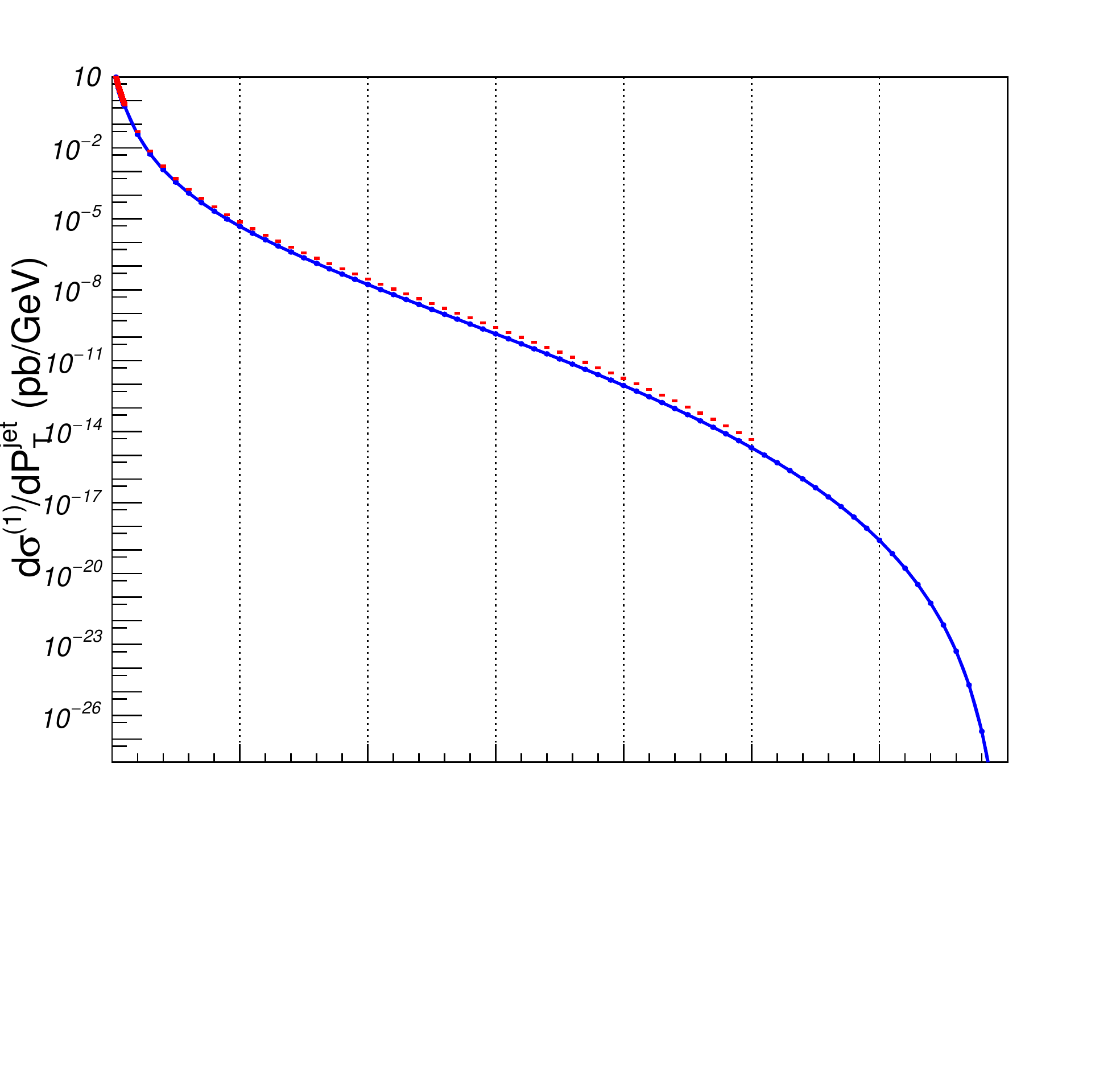}
\includegraphics[width=7.7cm]{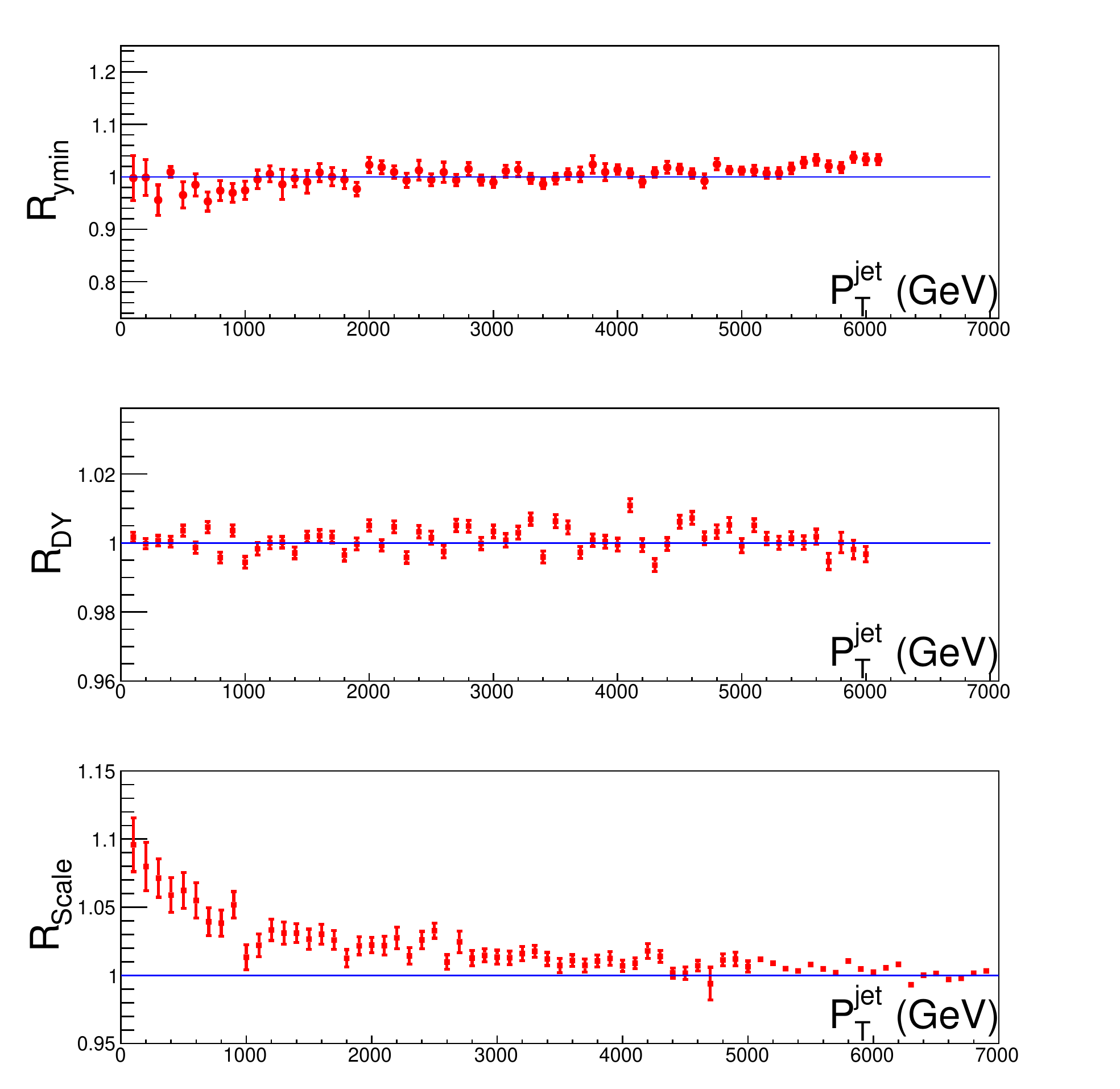}
\end{center}
\caption{
The transverse momentum distribution of the jet where
$|\eta_{\mbox{\tiny jet}}|<3$, $\mu_R=\mu_F=\sqrt{\hat s}$ 
and the parton density function CT14nlo is shown on the left. The blue line is the LO 
prediction and the red line is the NLO prediction.
The ratio of the distribution for $y_{min}=10^{-4}$ over $y_{min}=10^{-3}$
is given in the top right figure, while the middle right figure
shows the ratio of the FBPS MC over the original DYRAD MC.
The bottom right figure shows the ratio of two different 
renormalization/factorization scale choices. See text for
further details and discussion.
}
\label{dispt}
\end{figure}
\begin{figure}[t]
\begin{center}
\includegraphics[width=7.5cm]{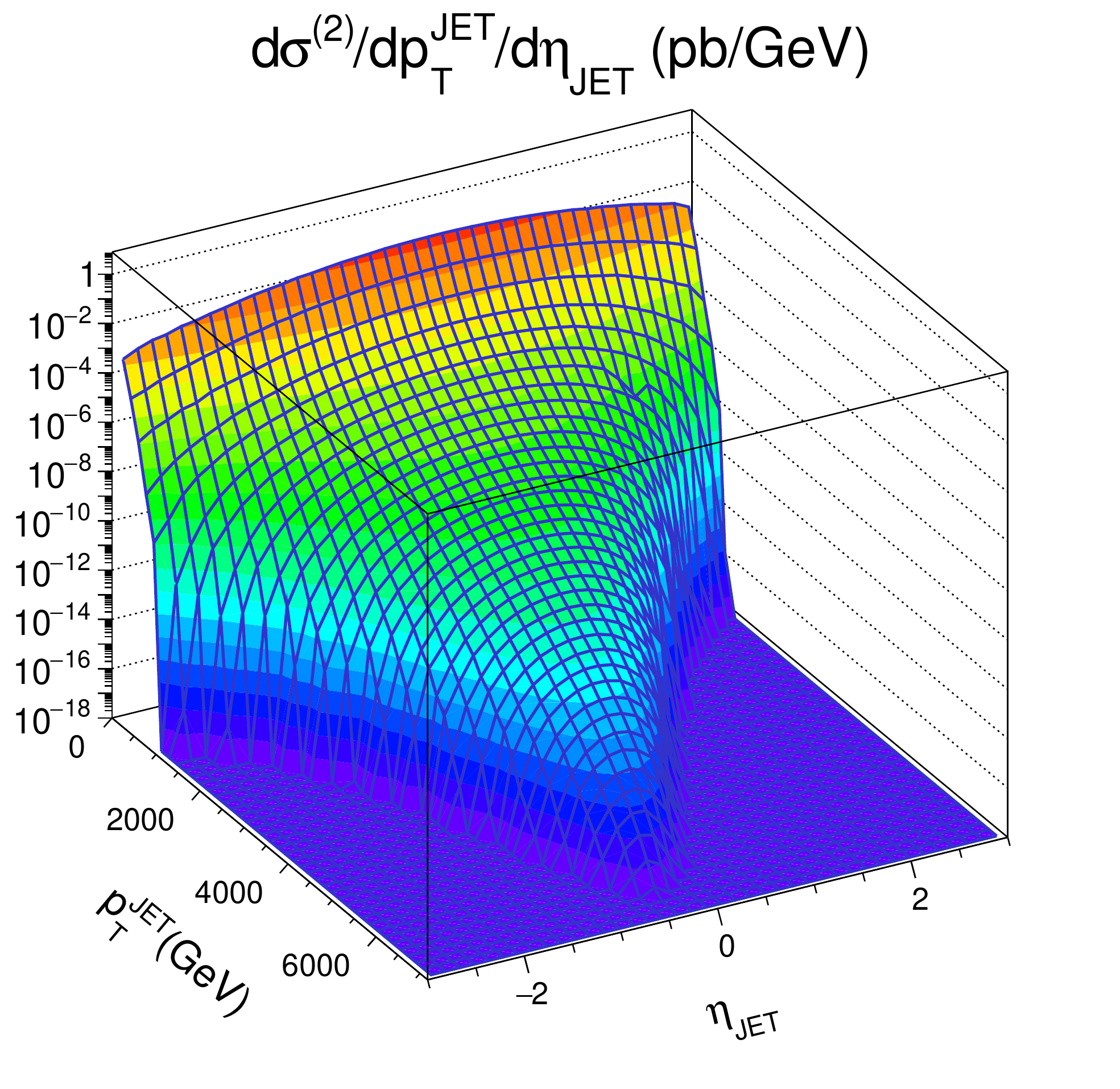}
\includegraphics[width=7.5cm]{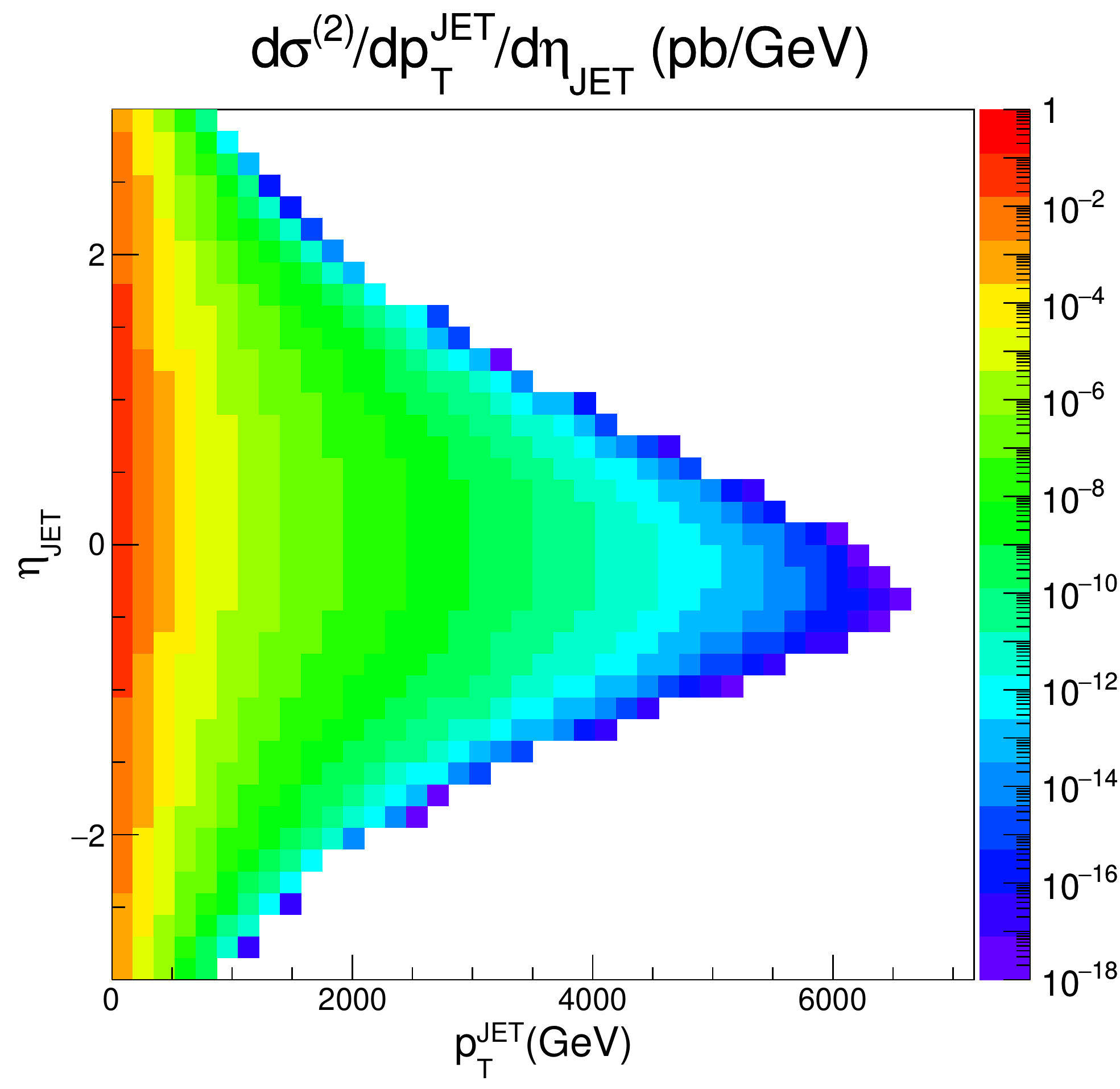}
\includegraphics[width=7.5cm]{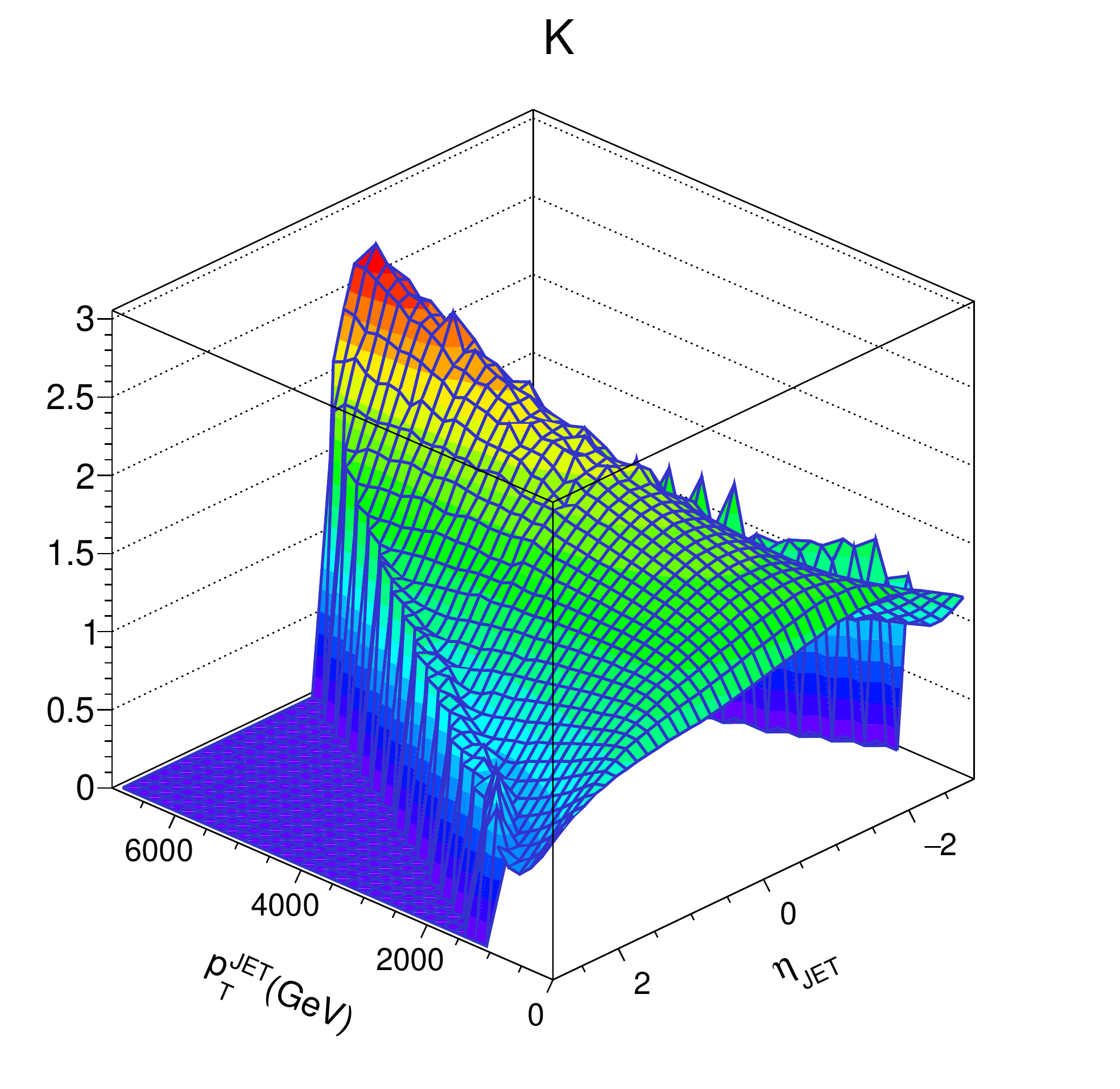}
\includegraphics[width=7.5cm]{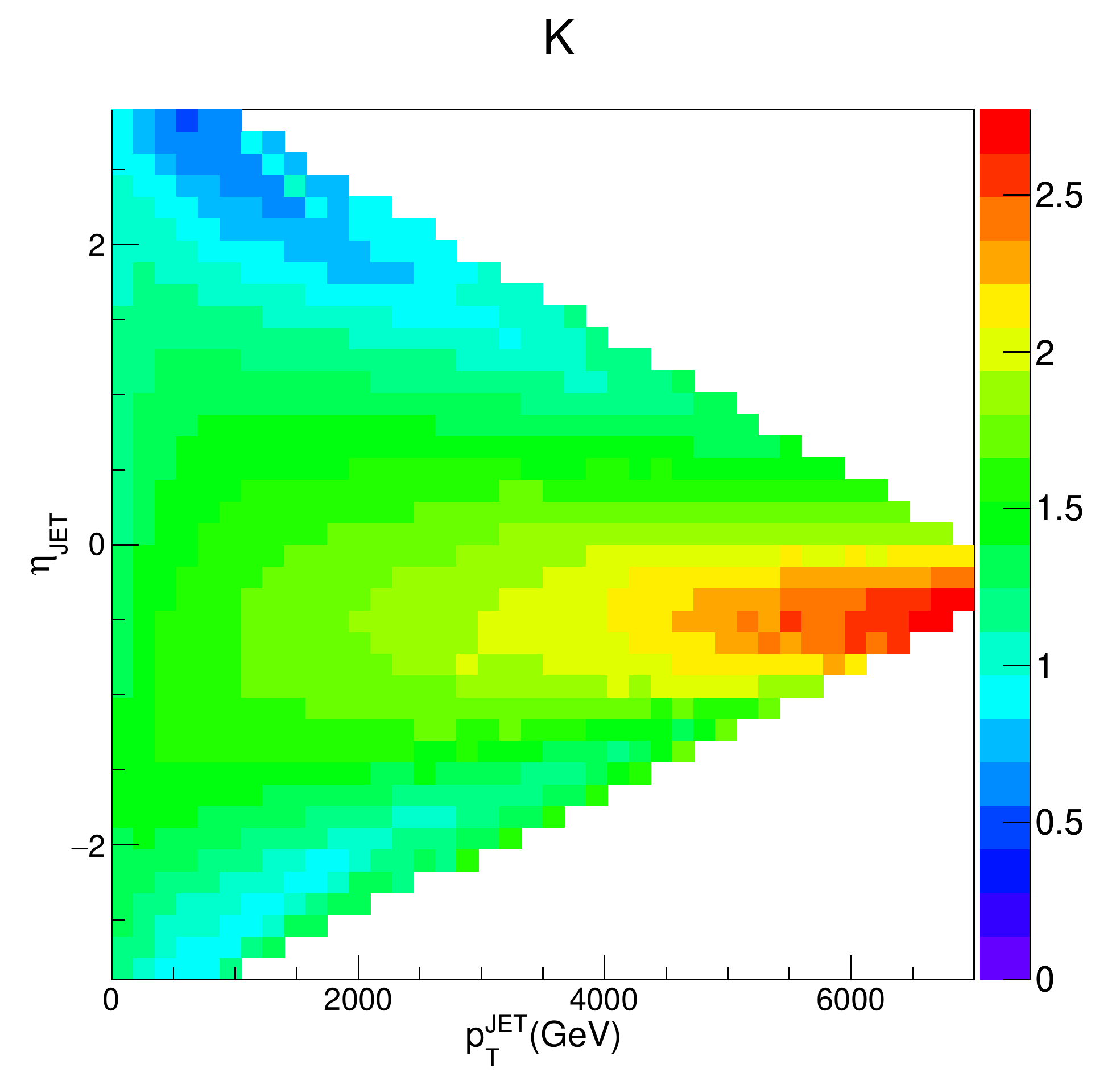}
\end{center}
\caption{
The 2-dimensional differential cross section (top row) and corresponding $K$-factor (bottom row)
$d\sigma_{\mbox{\tiny NLO}}^{(2)}/d p_T^{\mbox{\tiny JET}} d\eta_{\mbox{\tiny JET}}$. The
rapidity of the $Z$ boson is fixed to $\eta_Z=0.51$.
}
\label{dis-2D}
\end{figure}

Originally a theoretical study of a particular distribution at NLO
consisted of a calculation of the observable by hand with the aid of a non-MC computer
program. The constructed program was dedicated to the calculation of 
that particular observable/distribution at parton level.
There was no parton level MC approach or binning, instead
the program calculating the differential cross section was a mixture 
of analytic parts and some minor numerical integrations. It
calculated the value of the differential cross section for a specified input value of 
the observable (see e.g. refs.~\cite{Nason:1989zy,Ellis:1990ek}). 
The drawback of such an approach is that it is rigid,i.e.,the
definition of the observable is unalterable. Even the smallest change 
in the observable would
necessitate a new calculation. Also in order to be able to perform the calculation,
certain simplifications had to be made with respect to the experimental definition
of the observable. The needed computer resources were small and in-line with what
was available during that time-frame.

These drawbacks and the continued rise in available computer resources eventually
led to the advent of parton level MC programs 
(see e.g. refs.~\cite{Baer:1990ra,Giele:1994gf}) where the user  numerically
defines the observable resulting in a great flexibility. Moreover, the parton level MC
program could calculate any observable/distribution.
It has the expected drawbacks of a MC approach and requires
a large amount of computer resources to obtain a reasonable integration uncertainty
on the observable. Also, more responsibility was required from the user to calculate
sensible, infrared safe observables. As the available computer power increased 
significantly over the last few decades this model was sustainable for
more complicated NLO parton level MC's.

Because of technological limitations, currently different compute models 
are becoming more prominent.
Multi-threading combined with the use of cloud distributed node farms seems
to become the most efficient way to perform a computation. 
The need to calculate the differential cross section to ever more
precision and for more and more complex observables only has accelerated
this trend. The use  of brute force MC integration of the partonic
amplitudes using a simple phase space generator linked to an adaptive
MC like {\tt VEGAS} start to strain the available computer resources more
and more.
Because we could rely on ever more powerful CPU's in the last decades,
numerical phase space integrations never evolved beyond what was used in 
parton level MC like~\cite{Giele:1993dj,Giele:1994gf}. In contrast,
the techniques and methods to calculate scattering amplitudes 
dramatically increased.

Here we apply the FBPS approach to combine the advantages of the original,
more analytic approach and the parton MC approach within a framework of using
multi-threading and cloud computing.
To calculate distributions at parton level
we can fully exploit the advantages of the FBPS
approach. The distributions can be calculated on a finite grid of specific
values of the observables without any binning. 
For the calculation of the differential
cross section, one can push each grid point to a different cloud node.
This means we get precise predictions of the differential cross section
for all chosen values of the observables in matter of minutes, independent of
how many grid points we chose. At each value of the observable we can require a
specific accuracy. Populating the entire range of possible values of the
observable with the same relative integration uncertainty is trivial.  
We will use the Open Science Grid~\cite{Pordes:2007zz,Sfiligoi:2010zz}
for our calculations.

As a first example we explore in Fig.~\ref{dispt} the
transverse momentum distribution $p_T^{\mbox{\tiny JET}}$
of the jet. Given our inclusive partonic jet algorithm which clusters
all partonic particles into the jet, the transverse momentum
of the jet is almost identical to the transverse
momentum distribution of the $Z$-boson, was it not for the rapidity cut on the jet.
Because we fix the transverse momentum at a predetermined range of values,
$p_t^{\mbox{\tiny JET}}=1,2,\ldots,99,100,200,\ldots,6800,6900$ GeV,
integrate over the remaining LO 2-dimensional phase space and subsequently
calculate the $K$-factor by integrating over 3-dimensional \brem phase space,
we have excellent control over the integration accuracy at each $p_T^{\mbox{\tiny JET}}$
point selected. We chose the relative integration accuracy for this distribution to be 
no larger than 0.1\%. Additionally we can farm out each of the points to the grid.
The results for the LO and NLO differential cross sections are shown in Fig.~\ref{dispt},
together with the ratio of the NLO and LO differential cross section
given by the $K$-factor,
\beq
\frac{d\sigma^{(1)}_{\mbox{\tiny NLO}}}{d p_T^{\mbox{\tiny JET}}}=K(p_T)\times\frac{d\sigma^{(1)}_{\mbox{\tiny LO}}}{d p_T^{\mbox{\tiny JET}}}\ .
\eeq
The $K$-factor at both endpoints diverge as the \brem is inhibited approaching $p_T^{\mbox{\tiny JET}}\rightarrow 0$ GeV 
and  $p_T^{\mbox{\tiny JET}}\rightarrow 7000$ GeV. Away from the endpoints the $K$-factor gradually grows with increasing
transverse momentum 
between the negative divergence as $p_T^{\mbox{\tiny JET}}\rightarrow 0$ GeV and the
positive divergence as $p_T^{\mbox{\tiny JET}}\rightarrow 7000$ GeV.
Because we do not need to bin event weights from randomly generated
phase space point and can fix the $p_T^{\mbox{\tiny JET}}$ value we have
excellent control over the precision of the prediction at any $p_T^{\mbox{\tiny JET}}$-value of the
jet. This is in sharp contrast to generic MC approaches where populating certain transverse momentum
bins is complicated and maintaining a good accuracy over the entirety of the distribution is time
expensive.

To validate slicing parameter independence we shown in Fig.~\ref{dispt} the ratio of the $p_T$-distribution
choosing the slicing parameter $10^{-4}$
over the choice of $10^{-3}$. As is shown the choice of the slicing parameter has no effect on the 
prediction and only affects the run time to reach the required integration uncertainty.

As a final check on the predictions we check the binned prediction of the original DYRAD MC to
the FBPS version. To obtain the FBPS binned prediction we calculate
\beq
\left.\frac{\Delta\sigma}{\Delta p_T^{\mbox{\tiny JET}}}\right\rfloor_{p_T^{\mbox{\tiny MIN}}<p_T^{\mbox{\tiny JET}}<p_T^{\mbox{\tiny MAX}}}=
\frac{\sigma\left(p_T^{\mbox{\tiny JET}}>p_T^{\mbox{\tiny MIN}})-\sigma(p_T^{\mbox{\tiny JET}}>p_T^{\mbox{\tiny MAX}}\right)}{p_T^{\mbox{\tiny MAX}}-p_T^{\mbox{\tiny MIN}}}\ .
\eeq
As shown, the FBPS MC agrees well with the traditional parton level MC generator.

A final consideration is the renormalization/factorization scale choice. We have chosen the scale
to be the partonic center-of-mass energy $\sqrt{\hat s}$. This is in principle an incorrect choice
as we should construct the scale from the kinematics of the observables, i.e. the Born kinematics.
Such a choice would make the scale and hence $\alpha_S$ a constant when integrating over the
\brem phase space.
Constructing the Born equivalent of the partonic center-of-mass energy is readily done
\beqa
\hat s^{\mbox{\tiny LO}} &=& x_1^{\mbox{\tiny LO}}\times x_2^{\mbox{\tiny LO}}\times S\nonumber \\
\sqrt{S}\times x_{1/2}^{\mbox{\tiny LO}} &=& 
p_T^{\mbox{\tiny JET}}\times e^{\pm\eta_{\mbox{\tiny JET}}}+\sqrt{(p_T^V)^2+M_V^2}\times e^{\pm\eta_V}\ .
\label{scales}\eeqa 
Note that $\hat s>\hat s^{\mbox{\tiny LO}}$ for any Born configuration.  
In Fig.~\ref{dispt} we show the ratio of the NLO differential cross section using the two scales
and see that the difference is inconsequential. Although the LO partonic invariant mass gives
a slightly flatter $K$-factor we will use the
more intuitive choice of $\sqrt{\hat s}$ for the remainder of this section.

A 2-dimensional distribution can be calculated efficiently by extending the grid 
and by utilizing the cloud since the computational time is the same as for the 1-dimensional distribution.   
As an example we show in Fig.~\ref{dis-2D} the 2-dimensional NLO distribution and $K$-factor
\beq
\left.\frac{d\sigma_{\mbox{\tiny NLO}}^{(2)}}{d p_T^{\mbox{\tiny JET}} d\eta_{\mbox{\tiny JET}}}\right\rfloor_{\eta_Z=0.51}
=K\left(p_T^{\mbox{\tiny JET}}, \eta_{\mbox{\tiny JET}}\right)\left.\frac{d\sigma_{\mbox{\tiny LO}}^{(2)}}
{d p_T^{\mbox{\tiny JET}} d\eta_{\mbox{\tiny JET}}}\right\rfloor_{\eta_Z=0.51}
\eeq
where the $Z$-boson rapidity is fixed to a value $\eta_Z=0.51$ and transverse momentum and jet rapidity
grids are given by $p_T^{\mbox{\tiny JET}}=100, 200,\ldots, 6200$ GeV 
and $\eta_{\mbox{\tiny JET}}=-3.0,-2.9,-2.8,\ldots ,3.0$
resulting in a $61\times 61$ 2-dimension grid of points to be calculated for the distribution.
The results show a rich phenomenology with large $K$-factors near the kinematic boundaries.

\section{Connecting to the Experimental Observables}

\begin{figure}[t]
    \includegraphics[width=8.0cm]{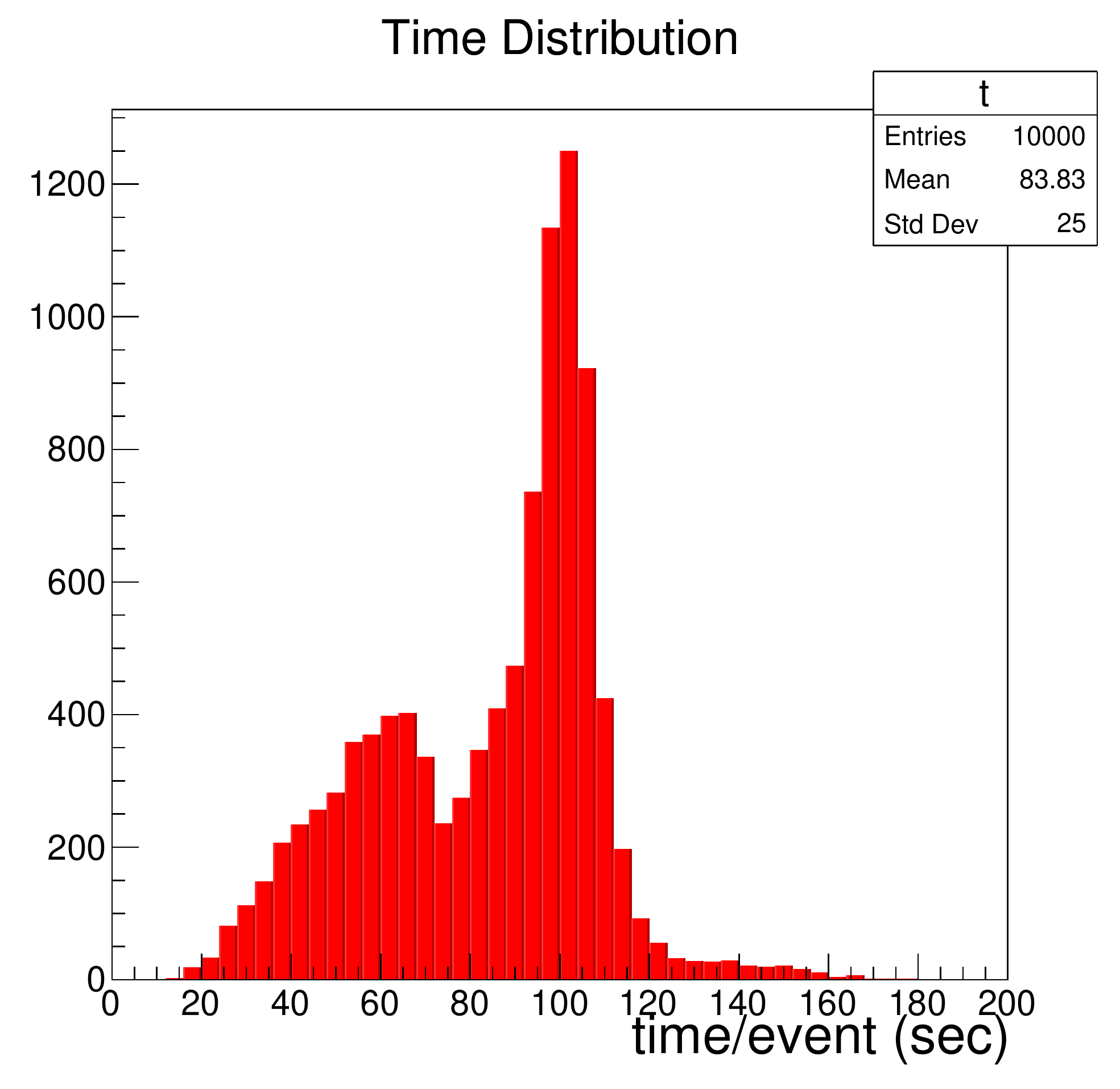}
    \includegraphics[width=8.0cm]{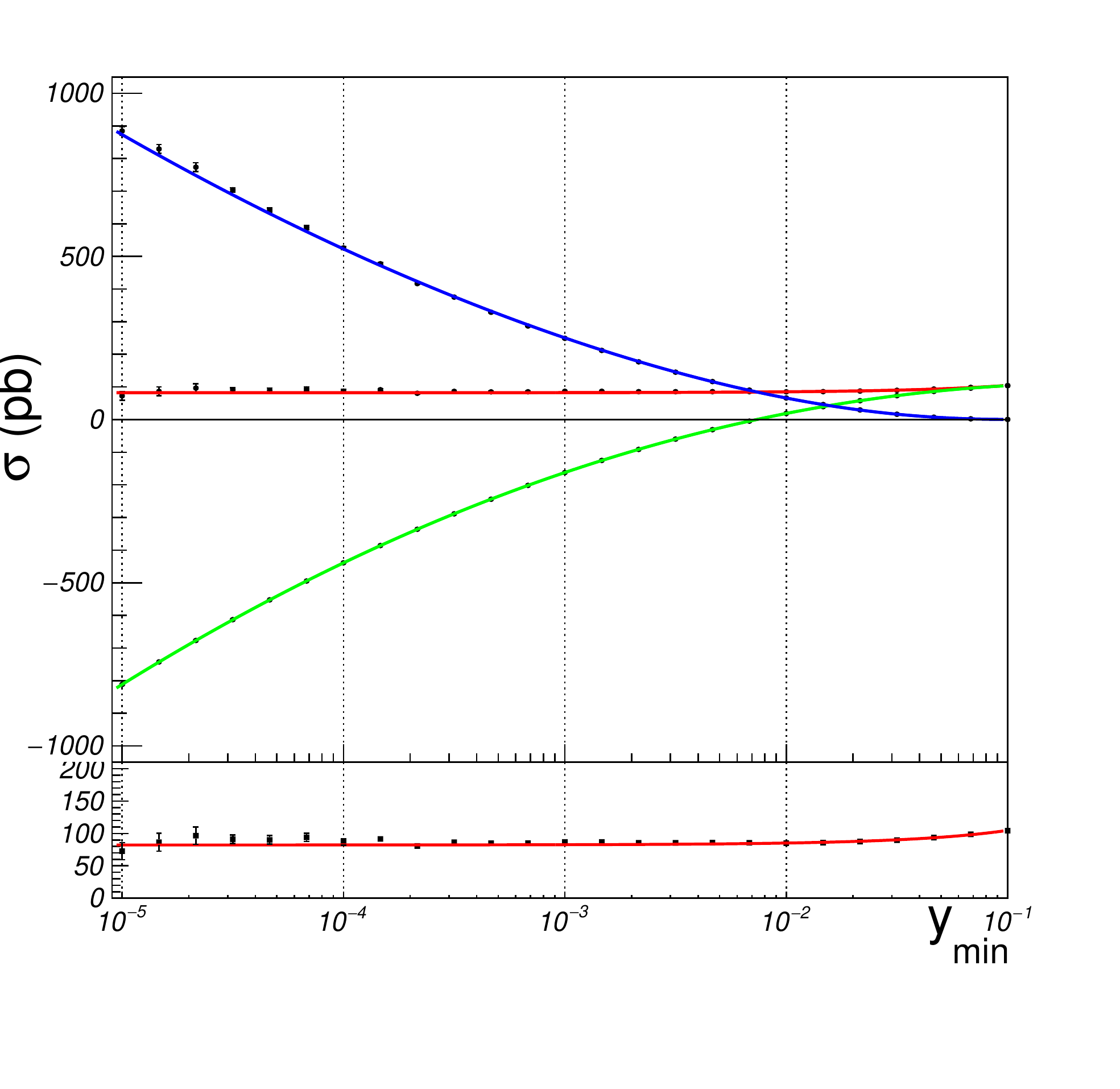}
\caption{The runtime per event on the Open Science Grid of 10,000 batches 
of 100 events for a total of 1,000,000 events
is shown in the left graph. The right graph gives the $y_{\mbox{\tiny min}}$- 
dependence of the exclusive 1-jet cross section
resulting in a integrated cross section of $82.2\pm 0.2$ pb.}
\label{Timing}
\end{figure}
\begin{figure}[htb]
  \begin{center}
    \includegraphics[width=7.7cm]{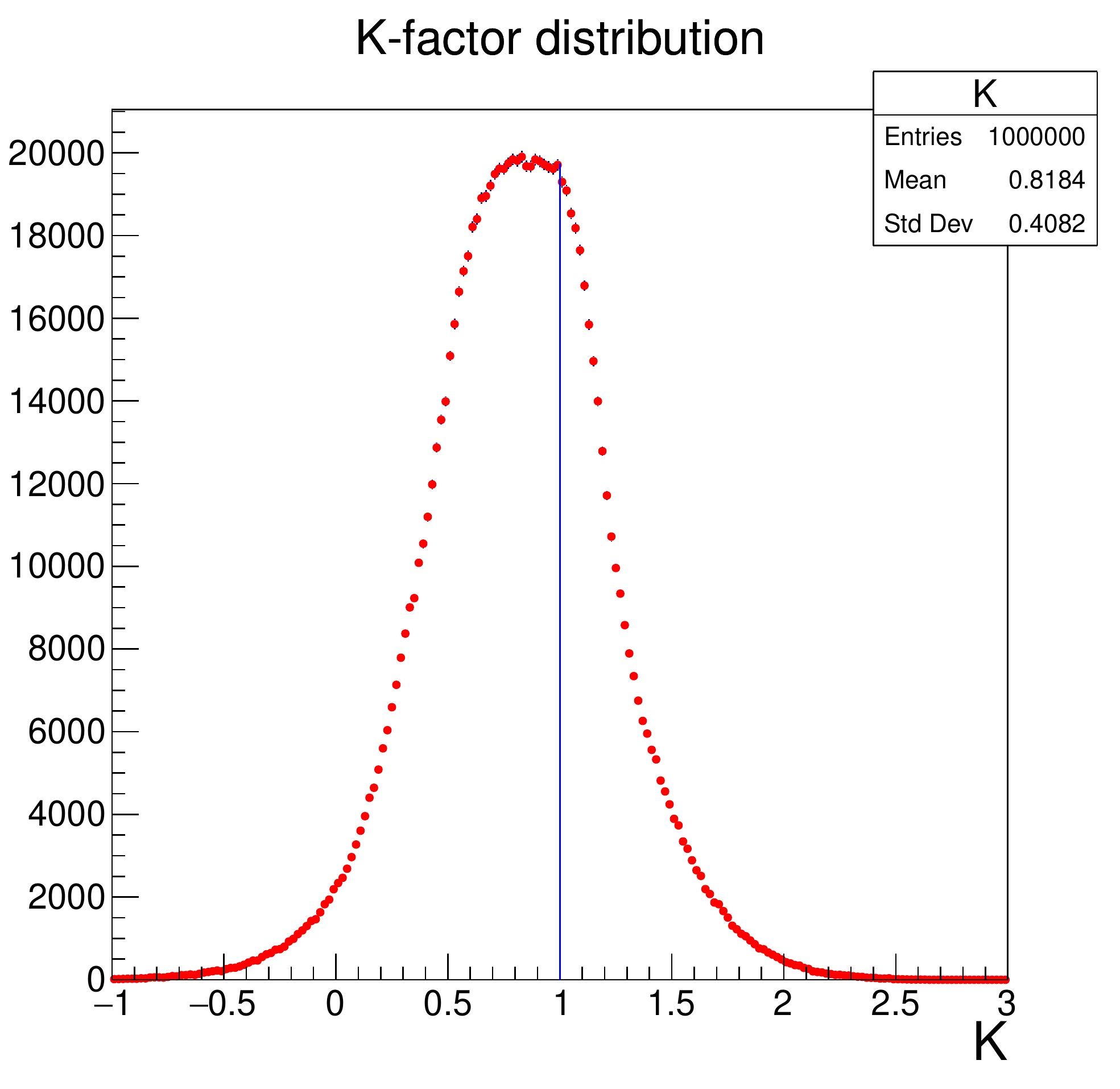}
    \includegraphics[width=7.7cm]{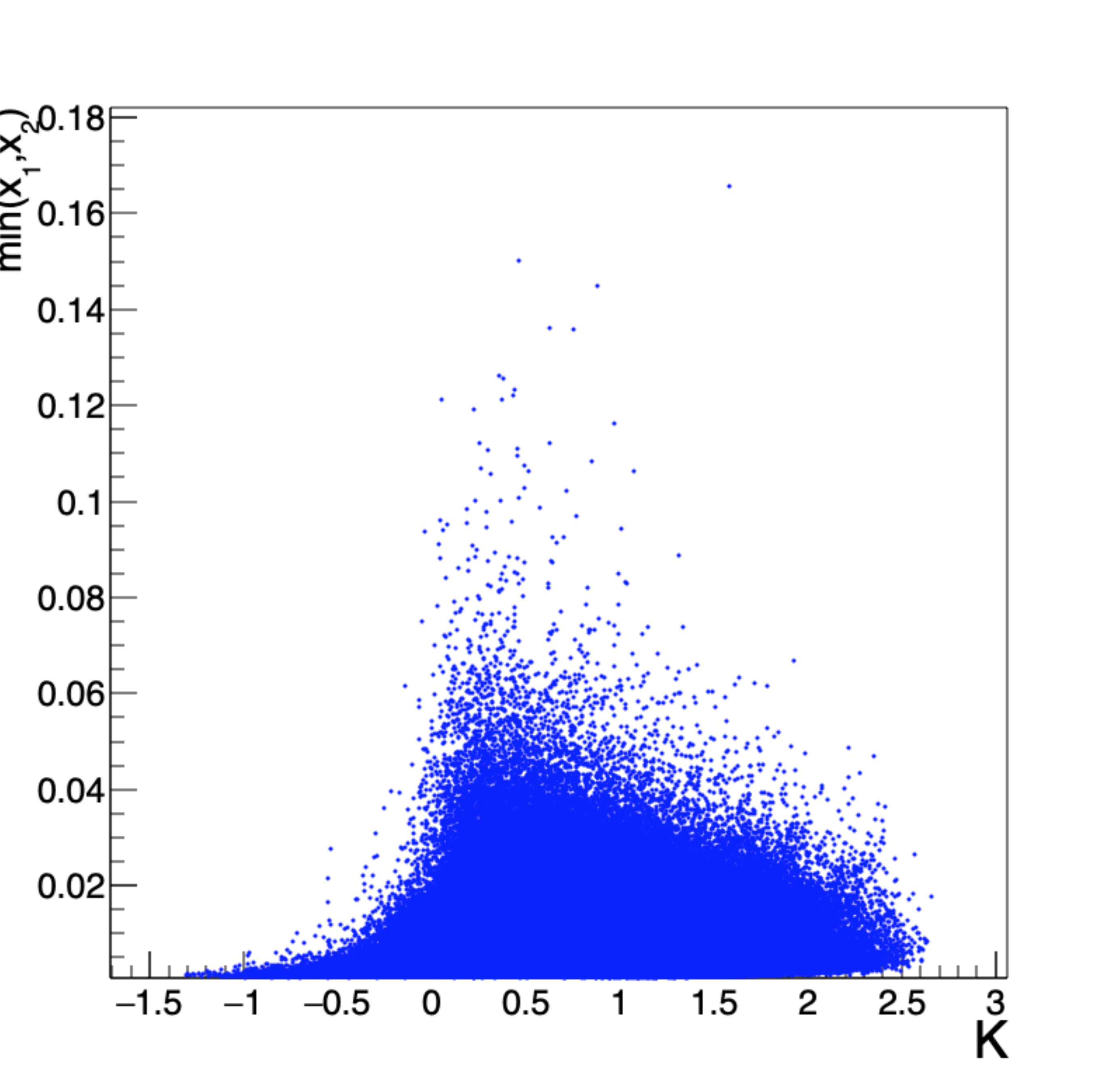}
\end{center}
\caption{The $K$-factor distribution of 1,000,000 unit-weight LO events on the left. 
The graph on the right shows the relation
between the $K$-factor and the smallest of the parton fractions in a scatter plot.}
\label{Kfactors}
\end{figure}

In the previous section we used the FBPS method on the parton level 
with the theoretical inspired inclusive jet algorithm.
While this is somewhat esoteric, a study using this framework exposes the underlying phenomenology
of the hard scattering process of interest to explore and understand. This contains all information
of the physics we can extract from the measurement and is useful to both theorists and 
phenemonologists. To connect to the physical process as observed by experimenters we need to
implement hadronization of the partonic state, including the recasting of the partonic jet algorithm 
into the hadronic jet algorithm with the main difference between the two 
being the beam remnant treatment which is absent for partonic jet
algorithms. One can consider an analytic approach as is outlined in Ref.~\cite{Dasgupta:2007wa}
which should give valuable insights into the hadronization process. However, to directly connect
to the experimental measurements one has to use a parton shower with its subsequent hadronization
and inclusion of the beam remnants. 

\begin{figure}[t]
  \begin{center}
   \includegraphics[width=7.7cm]{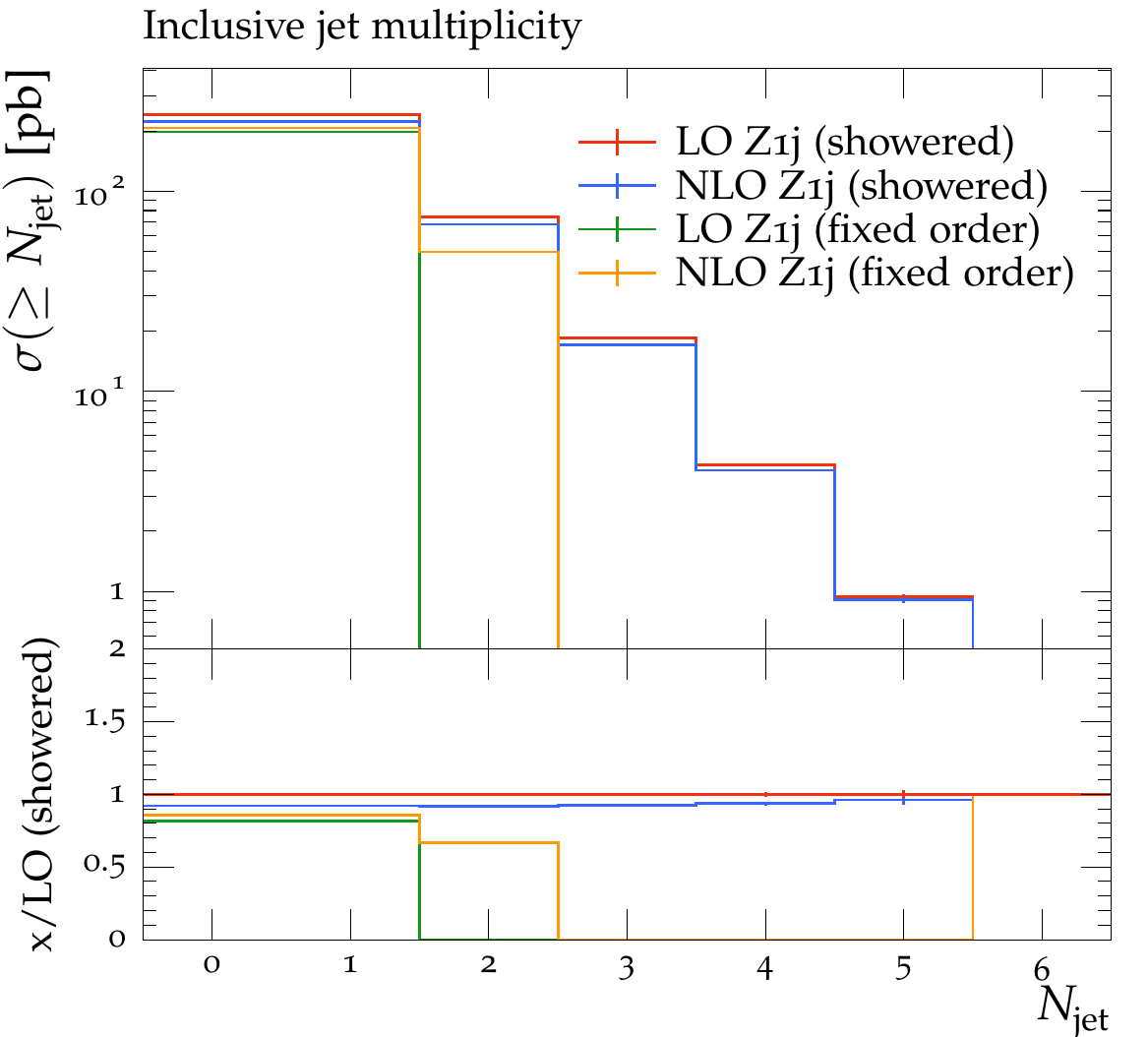}
   \includegraphics[width=7.7cm]{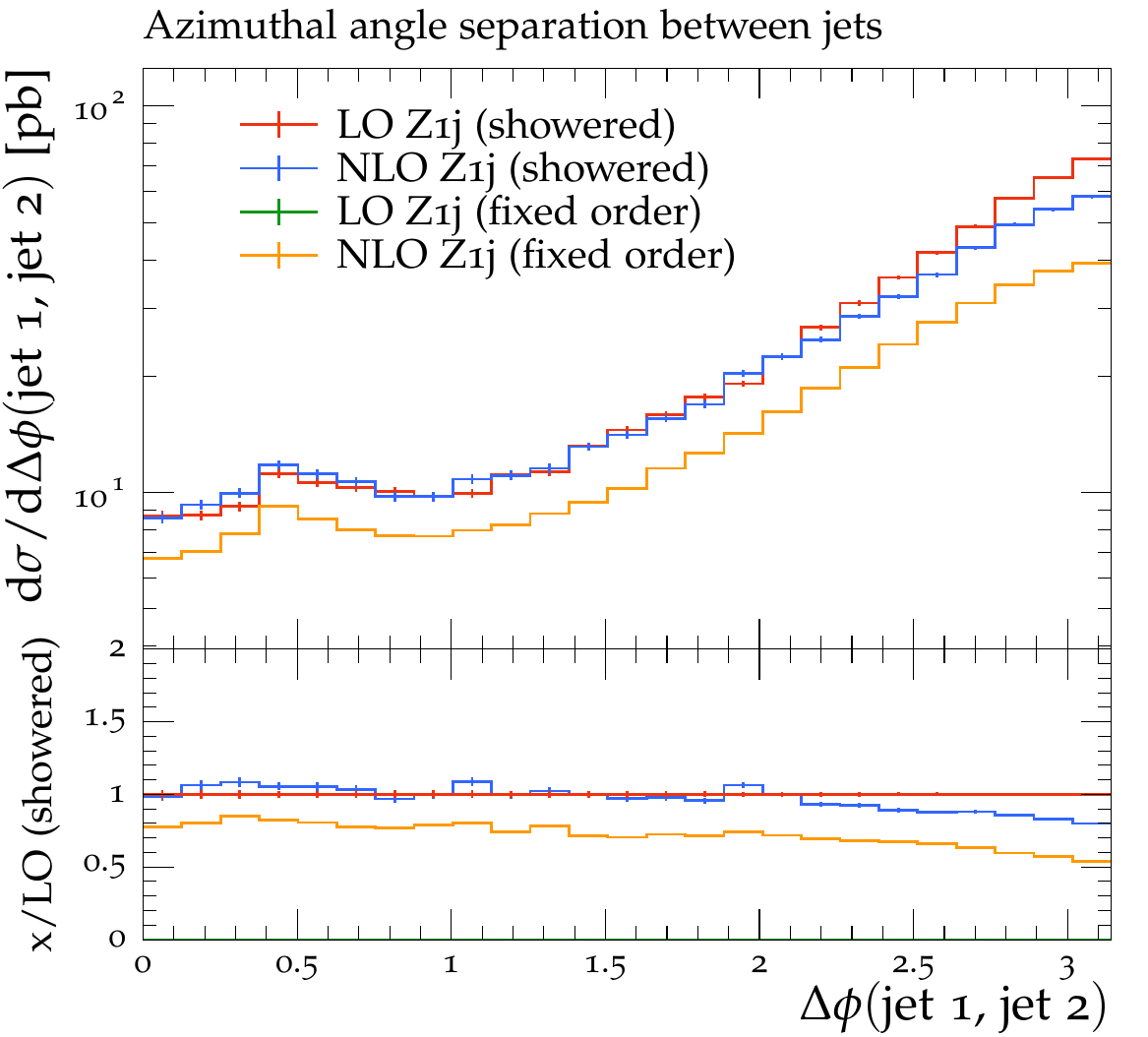}
   \includegraphics[width=7.7cm]{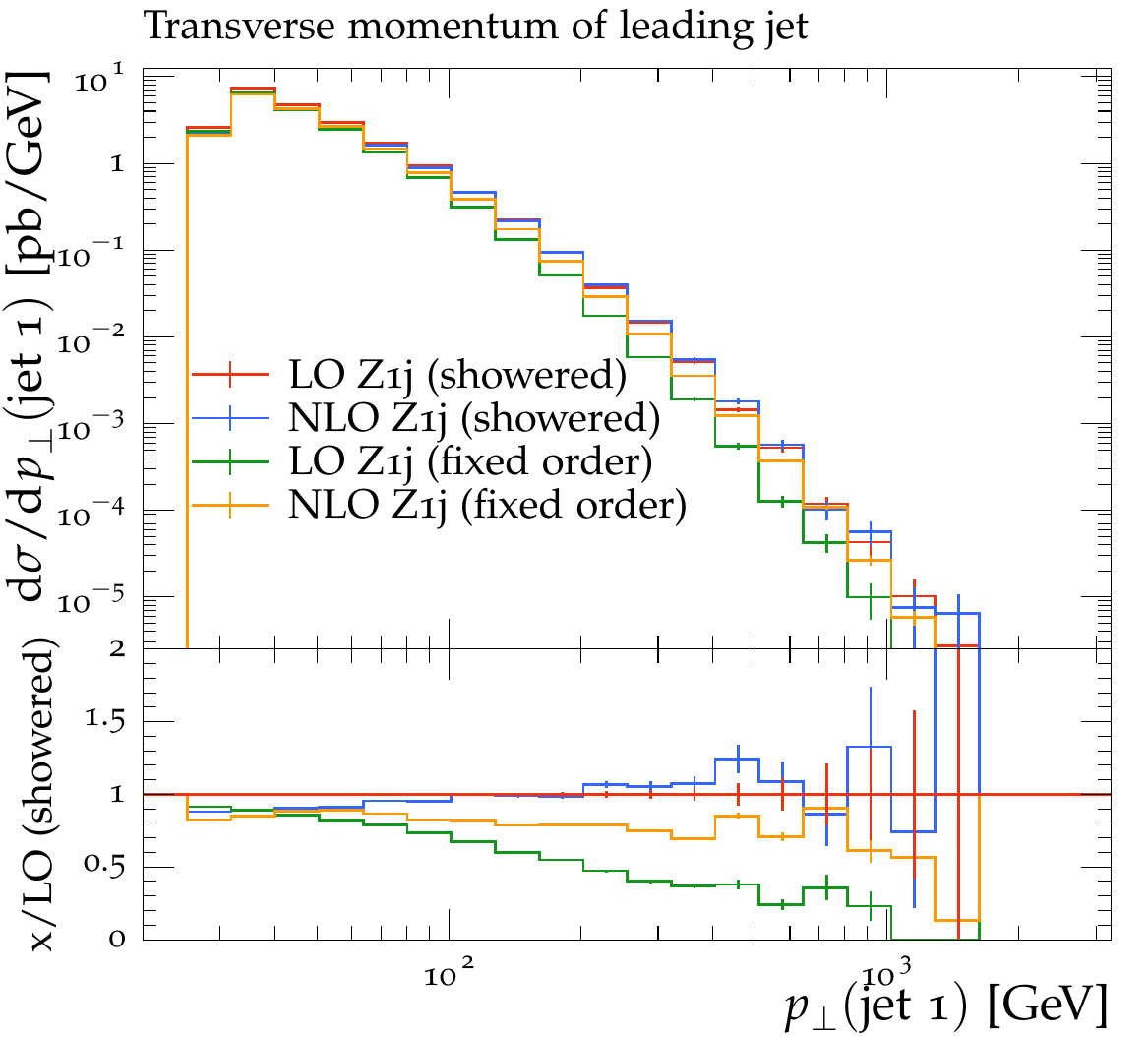}
   \includegraphics[width=7.7cm]{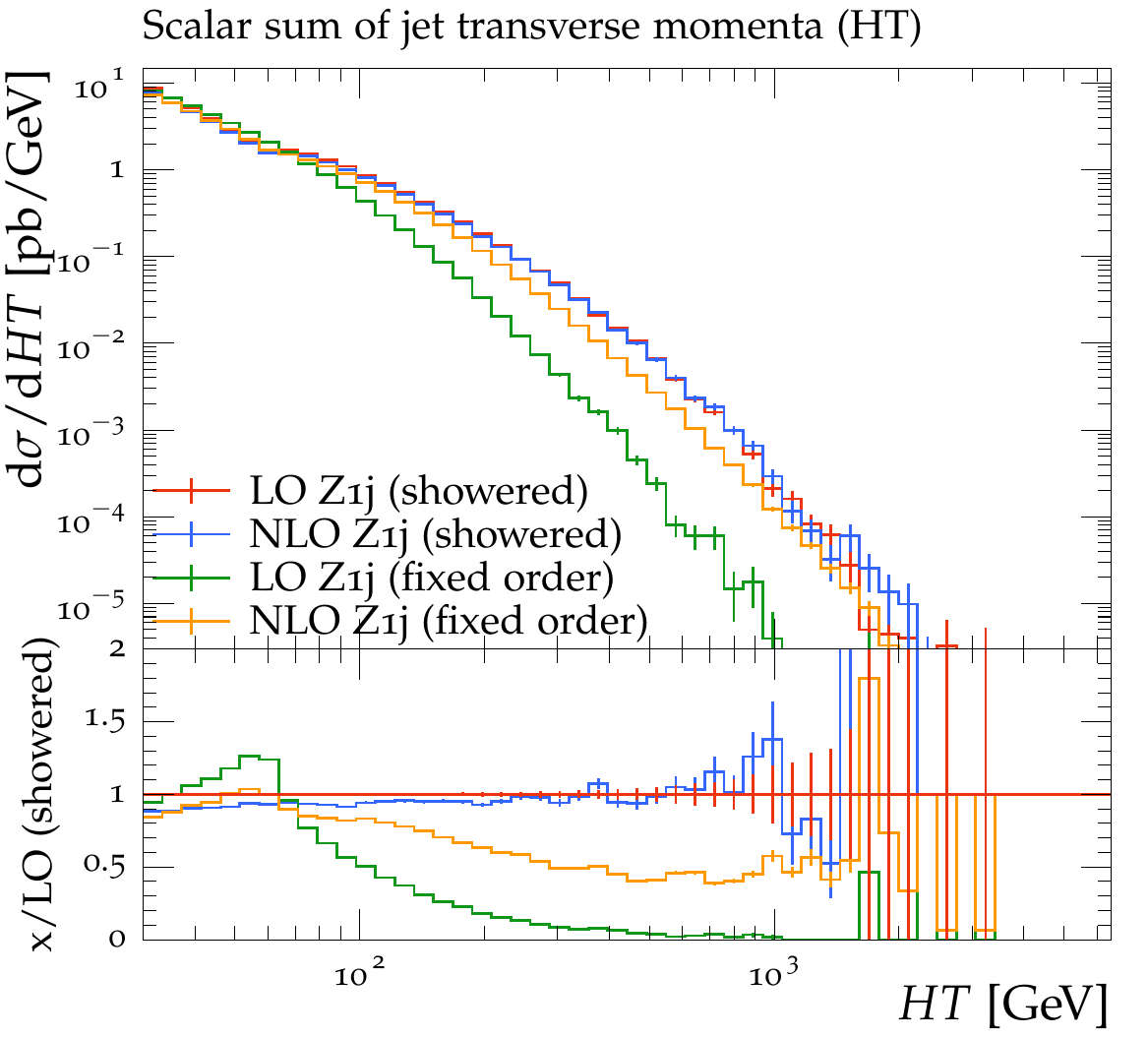}
   \includegraphics[width=7.7cm]{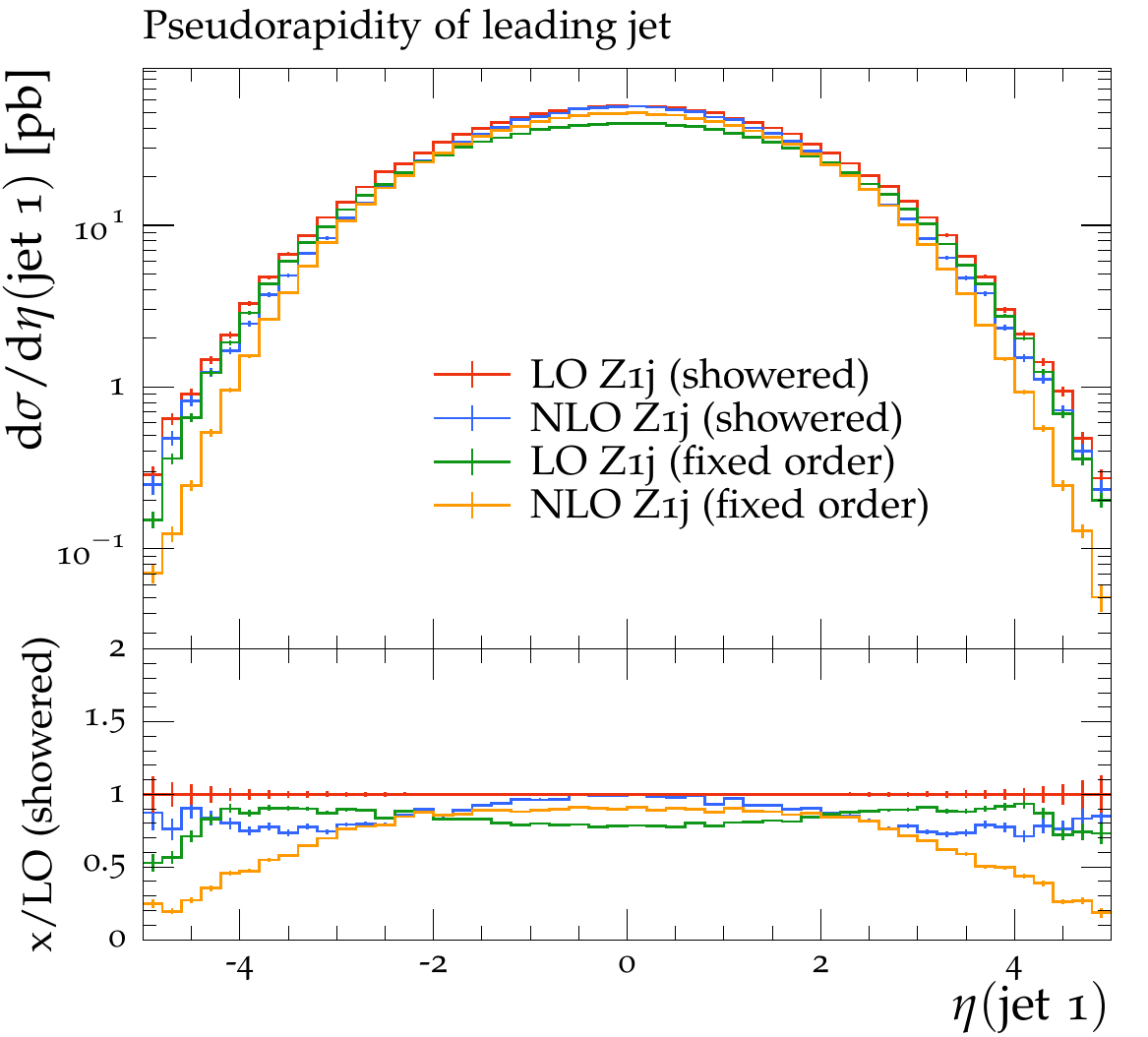}
   \includegraphics[width=7.7cm]{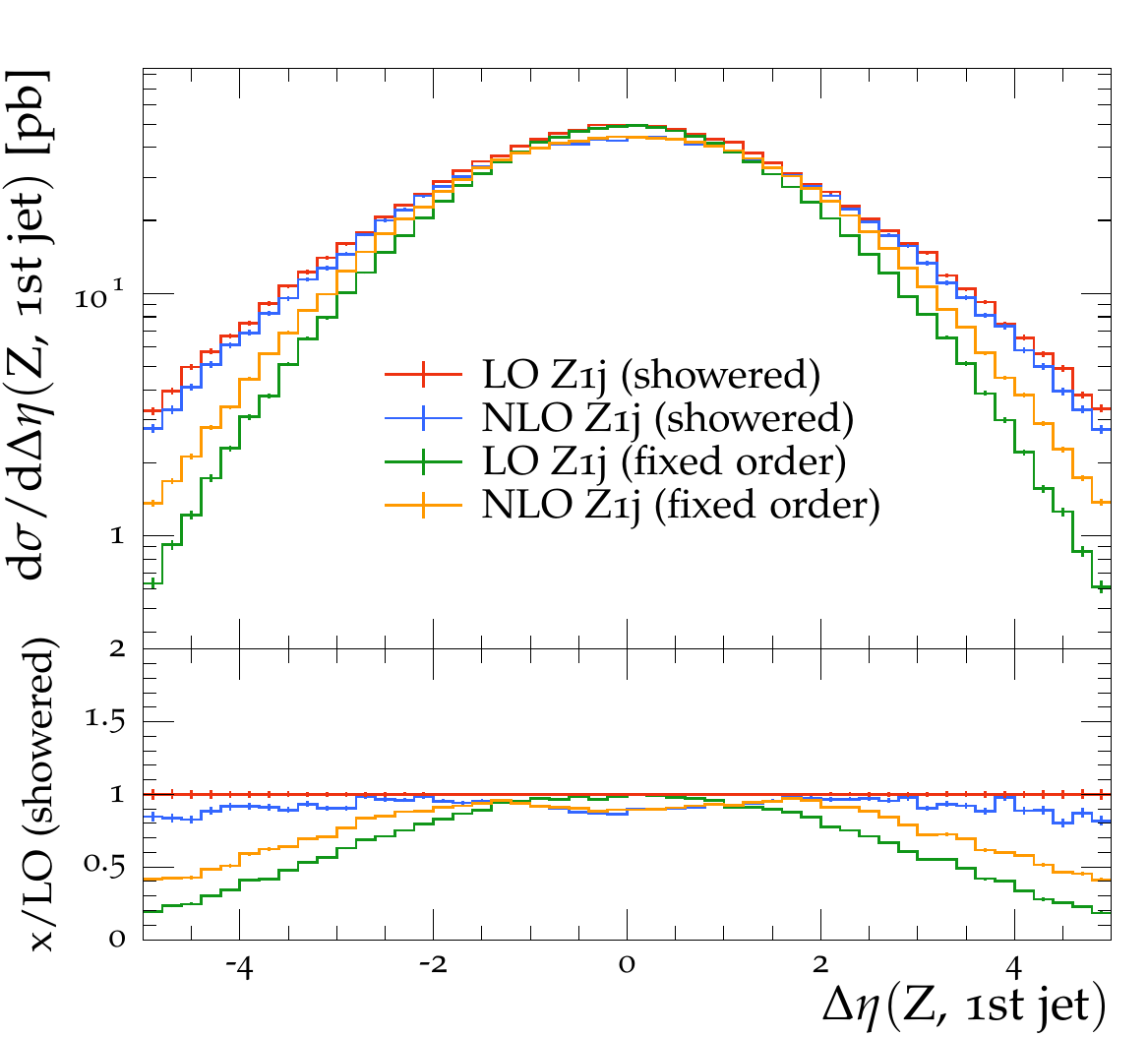}
  \end{center}
   \caption{Shown are the pQCD LO (LO Z1j (fixed order)) and NLO (NLO Z1j (fixed order)) predictions. 
 The inclusive showered results using {\tt VINCIA} originating from the pQCD events LO Z1j (showered) and 
 NLO Z1j (showered).  }
   \label{showered}
 \end{figure}

As described in Sec.~2 we will implement an inclusive partonic jet
showering approach for generating the hadronic
sample of $PP\rightarrow Z+(\geq) 1$ jets events. Before starting this analysis it
is informative to have a look at some generic properties relevant for the production
of large event samples. Both inclusive and exclusive jet algorithm methods start from
a sample of 1,000,000 unit-weighted LO $PP\rightarrow Z+1$ jet events which we generated using
Madgraph~\cite{Alwall:2011uj,Alwall:2014hca} for a collider energy of $\sqrt{S}=14$TeV. For each of these events we need to calculate the $K$-factor
which depends on the chosen jet algorithm.
In Fig.~\ref{Timing} we show the run times for calculating
the $K$-factors using the Open Science Grid. We submitted 10,000
jobs each containing 100 events to the cloud for a total of 1,000,000 calculated $K$-factors. 
As can be seen the evaluation time ranges over
quite a large range. The run-time depends on the properties of the specific node which can vary. 
The figure shows the average run-time per event, indicating that if we use 1,000,000
cloud nodes simultaneously the calculation would be done in minutes. Instead we used 100
events per cloud node, reducing the number of required nodes to 10,000. This means the overall
run-time is of order hours, still sufficiently fast for a quick turn-around. 
Also show in Fig.~\ref{Timing} is the slicing parameter $y_{\mbox{\tiny MIN}}$-dependence
of the total cross section using the exclusive partonic jet algorithm with the cuts and 
$K_T$-algorithm parameter taken from the ATLAS paper of Ref.~\cite{Aaboud:2017hbk}. 
The slicing parameter dependence of various observables 
using the inclusive partonic jet algorithm were already given in Sec.~3.

\begin{figure}[t]
  \begin{center}
   \includegraphics[width=7.7cm]{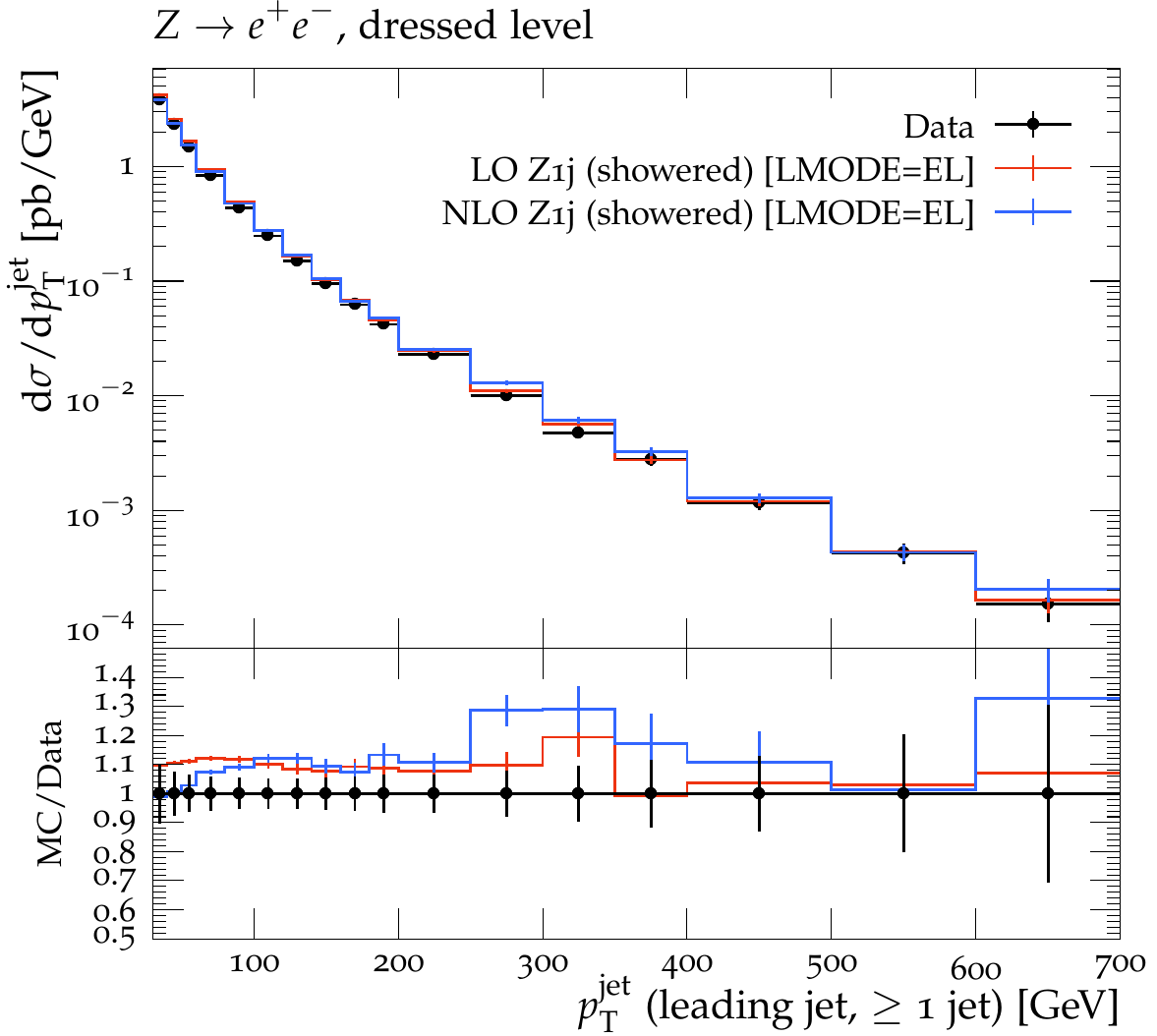}
   \includegraphics[width=7.7cm]{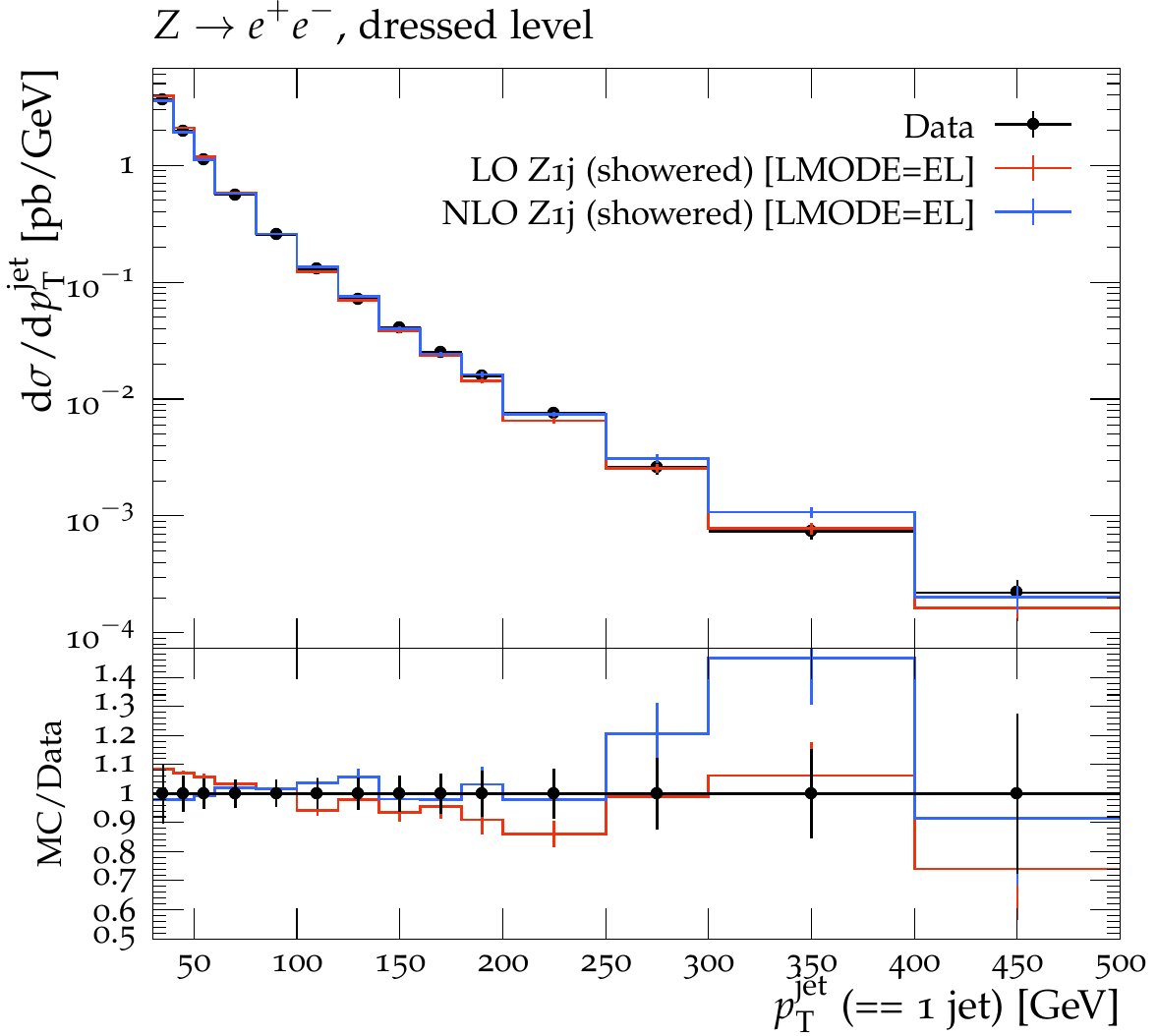}
  \end{center}
   \caption{
 The inclusive showered results using {\tt {\tt VINCIA}} originating from the pQCD events LO Z1j (showered) and 
 NLO Z1j (showered). Theory predictions are compared to the measured differential production cross sections from $3.16~{\rm fb}^{-1}$ of ATLAS data at the CERN 
 Large Hadron Collider for $Z$ boson
 in association of jets in proton-proton collision at $\sqrt{S}=13$~TeV\cite{Aaboud:2017hbk}. }
   \label{zjetsdata}
 \end{figure}

The sample of hadronic events is generated 
using the inclusive partonic jet algorithm. To calculate the $K$-factors for each
of the 1,000,000 unit-weight LO events we integrate over the full 3-dimensional
\brem phase space as the inclusive partonic jet algorithm does not depend on
any parameters other than the number of final state clusters at which we stop clustering which
in our case is one cluster. To get some feeling for the $K$-factors we show in Fig.~\ref{Kfactors}
the distribution of $K$-factors. We see the mean $K$-factor is close to 1 as expected. We also
see the $K$-factors can become negative for a small number of events. This is to be expected and to
what extent this occurs depends on the cuts employed. To get some indication about the 
phase space regions which leads to negative $K$-factors we make a scatter plot using the 
$\min(x_1^{\mbox{\tiny LO}},x_2^{\mbox{\tiny LO}})$. We see the negative $K$-factors are connected
with small parton fraction events. These events are typically low transverse momentum, large
rapidity events. The fact that negative $K$-factors exist makes re-weighting the LO events 
with the $K$-factors the best procedure. Interpreting the $K$-factors as some probability
is not possible.

The showered final state sample is generated using the {\tt VINCIA} shower 
Monte Carlo~\cite{Giele:2007di,Giele:2011cb,Fischer:2016vfv}. The subsequent analysis of the events was 
performed using {\tt RIVET} \cite{Buckley:2010ar}. Jets are reconstructed using the Anti-$K_{T}$ jet algorithm 
with $R=0.4$ \cite{Cacciari:2008gp} using {\tt FASTJET} \cite{Cacciari:2011ma}. At least one jet with $p_{T}>30$ GeV was 
required in a rapidity range of $[-5,5]$. Leptons are required to have transverse momentum of $p_{T}>10$ GeV and 
pseudo-rapidity in the range of $[-3.5,3.5]$. Events where the invariant mass of a electron and position 
between $71$ GeV and $111$ GeV are accepted, i.e, a single $Z$-boson is required.
Besides the pQCD LO and NLO predictions we also show in Fig.~\ref{showered} the results by showering
the unit-weight LO events and applying the inclusive $K$-factor to obtain the showered NLO results.
When applying the $K$-factor to the showered LO event we explore the most
basic matching scheme. By applying the $K$-factor to the
LO events we integrate the \brem over the allowed phase space. That is, the clustered partons
have the same transverse momentum vector and rapidity as the originating LO jet. We use the
parton shower to refill the integrated out \brem phase space with partons. This is per definition
unitary as the shower is unitary. The entirety of the resulting 1-jet inclusive sample
is at pQCD order $\alpha_S^2$ enhanced by a Leading Log shower as the {\tt VINCIA} shower is ME corrected. 

In Fig.~\ref{zjetsdata}, we compare our showered LO and NLO $Z+1$ jet results 
against the measured differential cross section from 
$3.16~{\rm fb}^{-1}$ of ATLAS data at the CERN Large Hadron Collider experiment 
for the transverse momentum for the leading jet $p_{T}^{jet}$ 
for events with one or more jets (inclusive) and one jet (exclusive) \cite{Aaboud:2017hbk}. Note that we have made node
effort to tune {\tt VINCIA} or the renormalization and factorization scales involved in the calculation of the $K$-factor. 
In a future paper, we plan to perform a more careful comparision to all available experimental collider data such 
as the measured $Z$ boson plus jets production cross section for proton-proton collisions at $\sqrt{S}=8$ TeV~\cite{Aad:2019hga}.

Because of the jet recasting we can use the theoretical motivated inclusive partonic jet algorithm which 
quite naturally has no knowledge of hadronization and beam remnants. Such physics is introduced
by the shower MC in our approach. This recasts the jet from a partonic object to a fully hadronic object
including physics like out of cone radiation and initial state radiation while at the same time
preserving the pQCD $K$-factor of the event which is obtained by integrating 
over all possible decays of the event. Note that the hadronic sample requires
no additional resolution parameters beyond the LO $Z+1$-jet generation of unit weight events.
The only difference between LO shower (possibly detector corrected) sample and NLO 
is the $K$-factor reweighting. The $K$-factor is calculated at the parton level and can be
outsourced to a cloud/farm environment. This trivially implies unitarity preservation, with the
physical caveat of the $Z+0$-jet final state.

\section{Conclusions}

In this paper we revisited the fundamentals of the phase space integration 
needed to calculate higher order corrections for jet events at colliders.
This results in new methods to make predictions in pQCD.
The method is well suited for farm/cloud computing and gives access to
large resources of computer power through, for example, the Open Science Grid without the need of
gaining access to overburdened institutional computer resources. The reason goes back to the
matrix element approach origins of the FBPS method. The calculation is partitioned in individual
events, each can be outsourced to the cloud. Each partition involves the calculation of a single
Born event plus it virtual corrections and the integration of the \brem events of the partonic 
configurations over the phase space regions dictated by the partonic jet algorithm. The dimensionality
of the \brem phase space does not depend on the jet multiplicity, and is given by three times the 
number of \brem partons involved, giving just a 3-dimensional integration at NLO. The jet multiplicity
partitions phase space and sets up sectors to integrate over, i.e. one sector per jet. This does
not affect the dimensionality, but will require more MC events to accurately cover all the sectors with
sufficient number of events. Because of the low dimensionality at NLO, one can consider integration
methods not based on the Monte Carlo integration.

We also made the first steps towards a full fledged method to
construct Monte Carlo's in collider physics. It was shown that one can choose any partonic jet 
algorithm as long as it is infra-red safe and have summed four-vector sequential clustering. 
The partonic \brem contributing to a fixed (multi-) jet configuration is integrated over, resulting
in a $K$-factor to the Born amplitude. One
can choose to let a shower Monte Carlo repopulate this region of phase space, recasting the partonic
jet final state into a hadronic jet final state.
The method generalizes readily beyond NLO, as the FBPS generator is branching based and therefore
iterative like a parton shower.

The next steps are to extend the method to NNLO calculations and multi-jet final states. Furthermore
the matching methods to parton showers need to be studied beyond the initial steps in this paper
so comparisons to existing data can be made. 

\acknowledgments 

T.~F. acknowledges the kind support of the Fermi Theory Group. T.~F. thanks Robert Kolleck for conversations 
concerning the splitting of event files.  
W.~G. is supported by the DOE contract DE-AC02-07CH11359.

This research was performed 
using resources provided by the Open Science Grid, which is supported by the 
National Science Foundation and the U.S. Department of Energy's Office of Science.

\bibliographystyle{JHEP}
\bibliography{dyrad_fbps.bib}
\end{document}